\documentclass[prd,nofootinbib,preprint,superscriptaddress]{revtex4}
\pdfoutput=1
\usepackage[T1]{fontenc}
\usepackage{amsmath,amssymb}
\usepackage{epsfig}
\usepackage{graphicx}
\usepackage[usenames,dvipsnames]{color}
\usepackage{subfigure}
\usepackage{slashed}
\usepackage[colorlinks,citecolor=blue]{hyperref}
\usepackage{pdfpages}
\usepackage{color}

\begin{document}
\title{Connecting Light Dirac Neutrinos to a Multi-component Dark Matter Scenario in Gauged $B-L$ Model}
\author{Dibyendu Nanda}
\email{dibyendu.nanda@iitg.ac.in}
\affiliation{Department of Physics, Indian Institute of Technology Guwahati, Assam 781039, India}

\author{Debasish Borah}
\email{dborah@iitg.ac.in}
\affiliation{Department of Physics, Indian Institute of Technology Guwahati, Assam 781039, India}

\begin{abstract}
We propose a new gauged $B-L$ extension of the standard model where light neutrinos are of Dirac type, naturally acquiring sub-eV mass after electroweak symmetry breaking, without any additional global symmetries. This is realised by choosing a different $B-L$ charge for right handed neutrinos than the usual $-1$ so that the Dirac Yukawa coupling involves an additional neutrinophilic scalar doublet instead of the usual Higgs doublet. The model can be made anomaly free by considering four additional chiral fermions which give rise to two massive Dirac fermions by appropriate choice of singlet scalars. The choice of scalars not only helps in achieving the desired particle mass spectra via spontaneous symmetry breaking, but also leaves a remnant $Z_2 \times Z'_2$ symmetry to stabilise the two dark matter candidates. Apart from this interesting link between Dirac nature of light neutrinos and multi-component dark matter sector, we also find that the dark matter parameter space is constrained mostly by the cosmological upper limit on effective relativistic degrees of freedom $\Delta N_{\rm eff}$ which gets enhanced in this model due to the thermalisation of the light right handed neutrinos by virtue of their sizeable $B-L$ gauge interactions.
\end{abstract}

\maketitle

\section{Introduction}
In spite of convincing evidence for existence of light neutrino masses and their large mixing \cite{Tanabashi:2018oca}, the nature of light neutrinos is still unknown. While neutrino oscillation experiments (which have measured two mass squared differences and three mixing angles \cite{Esteban:2018azc}) are not sensitive to the nature of neutrino: Majorana or Dirac, experiments looking for neutrinoless double beta decay ($0\nu \beta \beta$), a promising signature of Majorana neutrinos, have not yet found any positive results. Though this does not necessarily rule out the Majorana nature, yet it is motivating to study the possibility of light Dirac neutrinos. With this motivation, several earlier works \cite{Babu:1988yq, Peltoniemi:1992ss, Chulia:2016ngi, Aranda:2013gga, Chen:2015jta, Ma:2015mjd, Reig:2016ewy, Wang:2016lve, Wang:2017mcy, Wang:2006jy, Gabriel:2006ns, Davidson:2009ha, Davidson:2010sf, Bonilla:2016zef, Farzan:2012sa, Bonilla:2016diq, Ma:2016mwh, Ma:2017kgb, Borah:2016lrl, Borah:2016zbd, Borah:2016hqn, Borah:2017leo, CentellesChulia:2017koy, Bonilla:2017ekt, Memenga:2013vc, Borah:2017dmk, CentellesChulia:2018gwr, CentellesChulia:2018bkz, Han:2018zcn, Borah:2018gjk, Borah:2018nvu, CentellesChulia:2019xky,Jana:2019mgj, Borah:2019bdi, Dasgupta:2019rmf, Correia:2019vbn, Ma:2019byo, Ma:2019iwj, Baek:2019wdn, Saad:2019bqf, Jana:2019mez} have discussed different ways of generating light Dirac neutrino masses by suitable extension of the standard model (SM) with new particles and symmetries.

Similarly, evidence from cosmology experiments like Planck suggests that a mysterious, non-luminous and nan-baryonic component of matter, known as dark matter (DM), gives rise to around $26\%$ of the present universe's energy density. In terms of density 
parameter $\Omega_{\rm DM}$ and $h = \text{Hubble Parameter}/(100 \;\text{km} ~\text{s}^{-1} 
\text{Mpc}^{-1})$, the present DM abundance is conventionally reported as \cite{Aghanim:2018eyx}:
$\Omega_{\text{DM}} h^2 = 0.120\pm 0.001$
at 68\% CL. Apart from cosmological evidences, there are several astrophysical evidences too strongly supporting the presence of DM \cite{Zwicky:1933gu,Rubin:1970zza,Clowe:2006eq}. In spite of such convincing astrophysical and cosmological evidences, the particle nature of DM is not yet known. Since none of the SM particles can be a realistic DM candidate, several beyond standard model (BSM) proposals have been floated in the last few decades, the most popular of them being known as the weakly interacting massive particle (WIMP) paradigm. In this WIMP paradigm, a DM candidate typically with electroweak (EW) scale mass and interaction rate similar to EW interactions can give rise to the correct DM relic abundance, a remarkable coincidence often referred to as the \textit{WIMP Miracle}~\cite{Kolb:1990vq}. However, the same electroweak type interactions could also give rise to DM-nucleon scattering at an observable rate which can, in principle, be observed at ongoing or future direct detection experiments like LUX \cite{Akerib:2016vxi}, PandaX-II \cite{Tan:2016zwf, Cui:2017nnn}, XENON1T \cite{Aprile:2017iyp, Aprile:2018dbl}, LZ~\cite{Akerib:2015cja}, XENONnT~\cite{Aprile:2015uzo}, DARWIN~\cite{Aalbers:2016jon} and PandaX-30T~\cite{Liu:2017drf}. However, there have been no observations of any DM signal yet in the experiments, putting stringent bounds on DM-nucleon scattering rates. Such null results have not only motivated the studies of beyond thermal WIMP paradigm but also a richer DM sector consisting of multiple DM components. Some recent proposals for multi-component DM can be found in \cite{Cao:2007fy, Zurek:2008qg, Chialva:2012rq, Heeck:2012bz, Biswas:2013nn, Bhattacharya:2013hva, Bian:2013wna, Bian:2014cja, Esch:2014jpa, Karam:2015jta, Karam:2016rsz, DiFranzo:2016uzc, Bhattacharya:2016ysw, DuttaBanik:2016jzv, Klasen:2016qux, Ghosh:2017fmr, Ahmed:2017dbb, Bhattacharya:2017fid, Ahmed:2017dbb, Borah:2017hgt, Bhattacharya:2018cqu, Bhattacharya:2018cgx, Aoki:2018gjf, DuttaBanik:2018emv, Barman:2018esi, YaserAyazi:2018lrv, Poulin:2018kap, Chakraborti:2018lso, Chakraborti:2018aae, Bernal:2018aon, Elahi:2019jeo, Borah:2019epq, Borah:2019aeq, Bhattacharya:2019fgs, Biswas:2019ygr,Bhattacharya:2019tqq} and references therein. As several of these works pointed out, apart from having larger allowed region of parameter space due to freedom of tuning relative DM abundances, such multi-component DM scenarios often offer complementary probes at experiments spanning out to different frontiers.

Motivated by growing interest in light Dirac neutrinos and multi-component DM scenario, here we propose a model where both of these can be accommodated naturally. Instead of choosing discrete symmetries to stabilise DM, here we consider gauged $B-L$ symmetry where $B$ and $L$ correspond to baryon and lepton numbers respectively. While gauged $B-L$ symmetric extension of the SM was proposed long ago \cite{Davidson:1978pm, Mohapatra:1980qe, Marshak:1979fm, Masiero:1982fi, Mohapatra:1982xz, Buchmuller:1991ce}, realising DM and light neutrino masses in the model require non-minimal field content or additional discrete symmetries. As far as we are aware of, there has been only one proposal so far to accommodate light Dirac neutrinos in a gauged $B-L$ model without any additional discrete or global symmetries. In \cite{Han:2018zcn}, authors considered such a possibility where light Dirac neutrino masses arise at radiative level. However, such radiative seesaw model requires an enlarged additional fermion content. In addition, since this model predicts single component DM, it suffers from stringent direct detection bound mentioned above. Here we show that light Dirac neutrino mass can be generated at tree level from a neutrinophillic Higgs doublet with very minimal particle content along with a two component fermion DM scenario. Two component fermion DM arises naturally as one possible solution to the anomaly cancellation conditions of the model. As we discuss below, such anomaly cancellation crucially depends upon the $B-L$ charge of right handed part of light Dirac neutrinos thereby connecting the origin of light neutrino mass with $B-L$ charges of DM as well as number of DM components. Apart from constraining the model from experimental bounds related to neutrino mass, collider searches, DM relic and DM-nucleon scattering rates, we also apply other bounds like perturbativity of different dimensionless couplings, bounded from below criteria of the scalar potential. More importantly, due to the Dirac nature of light neutrinos having additional gauge interactions, additional light degrees of freedom (right handed part of light Dirac neutrino) can be thermalised in the early universe, which is severely constrained from big bang nucleosynthesis (BBN) and cosmic microwave background (CMB) data. We show that the corresponding CMB-BBN bounds on additional light degrees of freedom constrain the DM parameter space more strongly compared to other relevant bounds. This is in sharp contrast with \cite{Han:2018zcn} where due to single component DM, the direct detection constraints remained strongest.

This paper is organised as follows. In section \ref{sec0}, we give a brief overview of gauged $B-L$ models with different solutions to anomaly conditions including the one we choose to discuss in details in this work. In section \ref{sec1}, we discuss our model in details followed by section \ref{sec:constraints} where we mention different existing constraints on model parameters. In section \ref{sec3}, we briefly discuss the relic abundance and direct detection of DM followed by discussion of our results in section \ref{sec4}. Finally we conclude in section \ref{sec5}.

\section{Gauged $B-L$ Symmetry}
\label{sec0}
As pointed out above, the $B-L$ gauge extension of the SM is a very natural and minimal possibility as the corresponding charges of all the SM fields under this new symmetry is well known. However, a $U(1)_{B-L}$ gauge symmetry with only the SM fermions is not anomaly free. This is because the triangle anomalies for both $U(1)^3_{B-L}$ and the mixed $U(1)_{B-L}-(\text{gravity})^2$ diagrams are non-zero. These triangle anomalies for the SM fermion content turns out to be
\begin{align}
\mathcal{A}_1 \left[ U(1)^3_{B-L} \right] = \mathcal{A}^{\text{SM}}_1 \left[ U(1)^3_{B-L} \right]=-3  \nonumber \\
\mathcal{A}_2 \left[(\text{gravity})^2 \times U(1)_{B-L} \right] = \mathcal{A}^{\text{SM}}_2 \left[ (\text{gravity})^2 \times U(1)_{B-L} \right]=-3
\end{align}
Interestingly, if three right handed neutrinos are added to the model, they contribute $\mathcal{A}^{\text{New}}_1 \left[ U(1)^3_{B-L} \right] = 3, \mathcal{A}^{\text{New}}_2 \left[ (\text{gravity})^2 \times U(1)_{B-L} \right] = 3$ leading to vanishing total of triangle anomalies. This is the most natural and economical $U(1)_{B-L}$ model where the fermion sector has three right handed neutrinos apart from the usual SM fermions and it has been known for a long time. However, there exists non-minimal ways of constructing anomaly free versions of $U(1)_{B-L}$ model. For example, it has been known for a few years that three right handed neutrinos with exotic $B-L$ charges $5, -4, -4$ can also give rise to vanishing triangle anomalies \cite{Montero:2007cd}. This model was also discussed recently in the context of neutrino mass \cite{Ma:2014qra, Ma:2015mjd} and DM \cite{Sanchez-Vega:2014rka, Sanchez-Vega:2015qva, Singirala:2017see, Nomura:2017vzp} by several groups. Another solution to anomaly conditions with irrational $B-L$ charges of new fermions was proposed by the authors of \cite{Wang:2015saa} where both DM and neutrino mass can have a common origin through radiative linear seesaw.

Very recently, another anomaly free $U(1)_{B-L}$ framework was proposed where the additional right handed fermions possess more exotic $B-L$ charges namely, $-4/3, -1/3, -2/3, -2/3$ \cite{Patra:2016ofq}. These four chiral fermions constitute two Dirac fermion mass eigenstates, the lighter of which becomes the DM candidate having either thermal \cite{Patra:2016ofq} or non-thermal origins \cite{Biswas:2016iyh}. The light neutrino mass in this model had its origin from a variant of type II seesaw mechanism and hence remained disconnected to the anomaly cancellation conditions. In a follow up work by the authors of \cite{Nanda:2017bmi}, these fermions with fractional charges were also responsible for generating light neutrino masses at one loop level. This particular anomaly cancellation solution with four chiral fermions having fractional $B-L$ charges was also studied in the context of inverse seesaw for light neutrino masses in \cite{Biswas:2018yus}. One can have even more exotic right handed fermions with $B-L$ charges $-17/3, 6, -10/3$ so that the triangle anomalies cancel \cite{Nanda:2017bmi}.

In the recent work on $U(1)_{B-L}$ gauge symmetry with two component DM \cite{Bernal:2018aon}, the authors considered two right handed neutrinos with $B-L$ charge -1 each so that the model still remains anomalous. The remaining anomalies were cancelled by four chiral fermions with fractional $B-L$ charges leading to two Dirac fermion mass eigenstates both of which are stable and hence DM candidates. The two right handed neutrinos with $B-L$ charge -1 take part in generating light neutrino masses via type I seesaw mechanism resulting in massless lightest neutrino. In another recent work \cite{Biswas:2019ygr}, while implementing type III seesaw in a gauged $U(1)_{B-L}$, it was found the triangle anomalies can be canceled by two component fermion dark matter.

In this work, we try to study the possibility of realising light Dirac neutrinos in a gauged $B-L$ model along with stable dark matter candidate without incorporating any additional discrete symmetries. In the minimal $U(1)_{B-L}$ model with three right handed neutrinos having $B-L$ charge $-1$ each, we can not have light Dirac neutrinos naturally as left and right handed neutrinos couple to the SM Higgs field. Even if we forbid the Majorana mass term of right handed neutrinos by suitable choice of singlet scalars, light Dirac neutrino mass of sub-eV order will require extreme fine tuning of Yukawa couplings of the order $\mathcal{O}(10^{-12})$. Even if we tolerate such extreme fine tunings, the model does not have a dark matter candidate. In earlier works \cite{Farzan:2012sa, Dasgupta:2019rmf}, radiative light Dirac neutrino mass and a stable DM candidate were shown to exist in a gauged $B-L$ model, but with several additional global symmetries. Here we consider different $U(1)_{B-L}$ charge (other than $-1$) for right handed neutrinos in order to prevent the Dirac Yukawa coupling with the SM Higgs.
If the right handed neutrinos are assigned $B-L$ charge $-2$, then the remaining anomalies are 
\begin{align}
\mathcal{A}_1 \left[ U(1)^3_{B-L} \right] =21  \nonumber \\
\mathcal{A}_2 \left[(\text{gravity})^2 \times U(1)_{B-L} \right] =3
\end{align}
These can be cancelled after introducing four chiral fermions $\chi_L, \chi_R, \psi_L, \psi_R$ having $B-L$ charges $13/9, 22/9, 1/9, 19/9$ respectively. This can be seen as 
\begin{align}
\mathcal{A}_1 \left[ U(1)^3_{B-L} \right] = \left( \frac{13}{9} \right)^3 +\left( -\frac{22}{9} \right)^3+\left( \frac{1}{9} \right)^3+\left( -\frac{19}{9} \right)^3=-21  \nonumber \\
\mathcal{A}_2 \left[(\text{gravity})^2 \times U(1)_{B-L} \right] =\left( \frac{13}{9} \right) +\left( -\frac{22}{9} \right)+\left( \frac{1}{9} \right)+\left( -\frac{19}{9} \right) = -3
\end{align}
It should be noted that the anomaly cancellation conditions we are solving here are same as the ones adopted in our earlier work \cite{Biswas:2019ygr} leading to type III seesaw for Majorana neutrinos. However, we are using a different solution to the anomaly conditions here from the earlier work. This is due to the fact that a singlet scalars of $B-L$ charges $1, 4$ were required to generate the masses of singlet chiral fermions $-7/5, -2/5, 6/5, -14/5$. However, in our model, we can not have a singlet scalar with $B-L$ charge $4$ as with the chosen $B-L$ charge of right handed neutrinos (2 in our model), such a singlet scalar will generate a Majorana mass after spontaneous symmetry breaking, making it impossible to realise light Dirac neutrino scenario.

In the next section, we will show that, if the additional scalar sector of the model is chosen appropriately, we can realise light Dirac neutrinos along with two component fermion DM naturally without incorporating any additional discrete symmetries.

\section{The Minimal Model with Light Dirac Neutrino and DM}
\label{sec1}
In this section we have discussed about our model in detail. We have extended the standard model gauge group with an additional local $\rm{U(1)_{B-L}}$ gauge group where B and L are denoting baryon and lepton numbers respectively of a particular field. Addition of this new gauge group introduces anomalies in the theory which can be canceled by including additional fermionic degrees of freedom in the theory. We have discussed the details of anomaly cancellation in the previous section. We have already mentioned that our main motivation is to generate the Dirac neutrino mass along with the stable DM candidate in the theory. Keeping this in mind, we have added three copies of right handed neutrinos with $B-L$ charge -2 each. They couple to lepton doublets via an additional Higgs doublet $\eta$ with $B-L$ charge -1 generating three Dirac neutrinos with sub-eV mass. However, the addition of these new fermionic fields will increase the anomaly which can be cancelled by adding four SM gauge singlet chiral fermions with fractional $B-L$ charges. We need at least two singlet scalars and one extra scalar doublet to generate the masses all new fermions. The fermion and scalar content of the model are shown in table \ref{tab:data1} and \ref{tab:data2} respectively. The necessity of the individual scalar fields will be discussed later. 
\begin{table}
\begin{center}
\begin{tabular}{|c|c|}
\hline
Particles & $SU(3)_c \times SU(2)_L \times U(1)_Y \times U(1)_{B-L} $   \\
\hline
$q_L=\begin{pmatrix}u_{L}\\
d_{L}\end{pmatrix}$ & $(3, 2, \frac{1}{6}, \frac{1}{3})$  \\
$u_R$ & $(3, 1, \frac{2}{3}, \frac{1}{3})$  \\
$d_R$ & $(3, 1, -\frac{1}{3}, \frac{1}{3})$  \\

$\ell_L=\begin{pmatrix}\nu_{L}\\
e_{L}\end{pmatrix}$ & $(1, 2, -\frac{1}{2}, -1)$  \\
$e_R$ & $(1, 1, -1, -1)$ \\
$\nu_R$ & $ (1, 1, 0, -2)$ \\
\hline
$\chi_L$ & $(1, 1, 0, \frac{13}{9})$ \\
$\chi_R$ & $(1, 1, 0, \frac{22}{9})$ \\
$\psi_L$ & $(1, 1, 0, \frac{1}{9})$ \\
$\psi_R$ & $(1, 1, 0, \frac{19}{9})$ \\
\hline

\end{tabular}
\end{center}
\caption{Fermion Content of the Model}
\label{tab:data1}
\end{table}
\begin{table}
\begin{center}
\begin{tabular}{|c|c|}
\hline
Particles & $SU(3)_c \times SU(2)_L \times U(1)_Y \times U(1)_{B-L} $   \\
\hline
$H=\begin{pmatrix}H^+\\
H^0\end{pmatrix}$ & $(1,2,\frac{1}{2},0)$  \\
$\eta=\begin{pmatrix}\eta^+\\
\eta^0\end{pmatrix}$ & $(1,2,\frac{1}{2}, -1)$  \\
\hline
$\phi_1$ & $(1, 1, 0, 1)$ \\
$\phi_2$ & $(1, 1, 0, 2)$ \\
\hline
\end{tabular}
\end{center}
\caption{Scalar content of the Minimal Model}
\label{tab:data2}
\end{table}

The Lagrangian of this model can be written as
\begin{eqnarray}
\mathcal{L}&=&\mathcal{L}_{SM} -\frac{1}{4} {B^{\prime}}_{\alpha \beta}
\,{B^{\prime}}^{\alpha \beta} + \mathcal{L}_{scalar} 
+ \mathcal{L}_{fermion}\;.
\label{LagT}
\end{eqnarray}

Here, $\mathcal{L}_{SM}$ represents the Lagrangian involving charged leptons, left handed neutrinos, quarks, gluons and electroweak gauge bosons. Second term denotes the kinetic term of new gauge boson ($Z_{BL}$) expressed in terms of field strength tensor ${B^\prime}^{\alpha\beta}=  \partial^{\alpha}Z_{BL}^{\beta}-\partial^{\beta}Z_{BL}^{\alpha}$. Please note that, in principle, the symmetry of the model allows a kinetic mixing term between $U(1)_Y$ of SM and $U(1)_{B-L}$ of the form $\frac{\epsilon}{2} B^{\alpha \beta} B^{\prime}_{\alpha \beta}$ where $B^{\alpha\beta}=  \partial^{\alpha}B^{\beta}-\partial^{\beta}B^{\alpha}$ and $\epsilon$ is the mixing parameter. Even if we turn off such mixing at tree level as we have done here, one can generate such mixing at one loop level since there are particles in the model which are charged under both $U(1)_Y$ and $U(1)_{B-L}$. Such one loop mixing can be approximated as $\epsilon \approx g_{\rm BL} g'/(16 \pi^2)$ \cite{Mambrini:2011dw}.  As we will see from final allowed parameter space in our numerical analysis, we have ${\rm g_{BL} \leq 0.2}$ for few TeV $B-L$ gauge boson mass and with such small values of ${\rm g_{BL}}$, the mixing parameter ${\rm \epsilon}$ will be of the order of ${\rm 10^{-3}}$ or smaller. Such small mixing has very little effect on the final allowed parameter space in our model, to be discussed in details in upcoming sections. Therefore, for simplicity, we ignore such kinetic mixing for the rest of our analysis.

The
gauge invariant scalar interactions described by $\mathcal{L}_{scalar}$ can be written as  
\begin{align}
\mathcal{L}_{scalar} &= \left({D_{H}}_{\mu} H \right)^\dagger
\left({D_{H}}^{\mu} H \right) + \left({D_{\eta}}_{\mu} \eta \right)^\dagger \left({D_{\eta}}^{\mu} \eta \right) + \sum_{i=1}^2 \left({D_{\phi_i}}_{\mu} \phi_i \right)^\dagger
\left({D_{\phi_i}}^{\mu}\,\phi_i \right)-\bigg \{-\mu^2_H \lvert H \rvert^2  \nonumber \\ 
& + \lambda_H \lvert H \rvert^4 + \left( \mu^2_{\eta} \lvert \eta \rvert^2 + \lambda_{\eta} \lvert \eta \rvert^4 \right)+\sum_{i=1,2} \left( -\mu^2_{\phi_i} \lvert \phi_i \rvert^2 + \lambda_{\phi_i} \lvert \phi_i \rvert^4 \right)  +\lambda_{H\eta} (\eta^{\dagger} \eta) (H^{\dagger} H)\nonumber \\
& + \lambda^{\prime}_{H\eta} (\eta^{\dagger} H) (H^{\dagger} \eta)  + \sum_{i=1,2}\lambda_{H\phi_i} (\phi^{\dagger}_i \phi_i) (H^{\dagger} H) + \left ( \lambda_{H \eta \phi} H^{\dagger} \eta \phi^*_1 \phi_2 +\text{h.c.} \right) \nonumber \\
&  + \left ( \mu_{H \eta} H^{\dagger} \eta \phi_1 +\text{h.c.} \right) + \sum_{i=1,2} \lambda_{\eta \phi_i} (\eta^{\dagger} \eta)(\phi^{\dagger}_i \phi_i) +\lambda_{\phi} (\phi^{\dagger}_1 \phi_1)(\phi^{\dagger}_2 \phi_2) \nonumber \\
& + \left (\mu_{\phi} \phi_1 \phi_1 \phi^{\dagger}_2 + \text{h.c.} \right) \bigg \}
\label{scalar:lag}
\end{align}
Where $\rm{{D_{H}}^{\mu}}$, $\rm{{D_{\eta}}^{\mu}}$ and $\rm{{D_{\phi}}^{\mu}}$ denote the covariant derivatives for the scalar doublets H, $\rm{\eta}$ and scalar singlets ${\rm\phi_i}$ respectively and can be written as 

\begin{eqnarray}
{D_{H}}_{\mu}\,H &=& \left(\partial_{\mu} + i\,\dfrac{g}{2}\,\sigma_a\,W^a_{\mu}
+ i\,\dfrac{g^\prime}{2}\,B_{\mu}\right)H \,, \nonumber \\
{D_{\eta}}_{\mu}\,\eta &=& \left(\partial_{\mu} + i\,\dfrac{g}{2}\,\sigma_a\,W^a_{\mu}
+ i\,\dfrac{g^\prime}{2}\,B_{\mu} + i\,g_{BL}\,n_{\eta}
{Z_{BL}}_{\mu}\right)\eta \,, \nonumber \\
{D_{\phi}}_{\mu}\,\phi_i &=& \left(\partial_{\mu} + i\,g_{BL}\,n_{\phi_i}
{Z_{BL}}_{\mu}\right)\phi_i\,.
\end{eqnarray}
where ${\rm g_{BL}}$ is the new gauge coupling and ${\rm n_{\eta}}$ and ${\rm n_{\phi_i}}$ are the charges under ${\rm U(1)_{B-L}}$ for ${\rm \eta}$ and ${\rm \phi_i}$ respectively. After both $B-L$ and electroweak gauge symmetries get spontaneously broken by the vacuum expectation value (VEV) of H and $\phi_i$s, the doublet and singlet scalars can be written as

\begin{eqnarray}
H=\begin{pmatrix}H^+\\
\dfrac{h^{\prime} + v + i z}{\sqrt{2}}\end{pmatrix}\,,\,\,\,\,\,\,
\eta=\begin{pmatrix}\eta^+\\
\dfrac{\eta_R^{\prime} + i \eta_I^\prime}{\sqrt{2}}\end{pmatrix}\,,\,\,\,\,\,\,
\phi_i = \dfrac{s^{\prime}_i +u_i+ A^{\prime}_i}{\sqrt{2}}\,\,\,\,(i=1,\,2)\,\,.
\label{H&phi_broken_phsae}
\end{eqnarray} 
From equation \eqref{H&phi_broken_phsae}, it is clear that the neutral component of the scalar doublet H and the scalar singlets ${\rm \phi_i}$ acquire non-zero VEV whereas the neutral component of $\rm{\eta}$ does not. This can be assured by suitably choosing the sign of bare mass squared term of $\eta$ field to be positive definite $(\mu^2_{\eta}>0)$. Even after the spontaneous symmetry breaking of $U(1)_{B-L}$, the effective bare mass squared term for $\eta$ can be assumed to be positive definite by appropriate choice of quartic couplings in the scalar potential. However, one crucial point to note here is that the neutral component of ${\rm\eta}$ will get a very tiny induced VEV after electroweak symmetry breaking because of the presence of trilinear term ${\rm H^{\dagger} \eta \phi_1}$ as well as the quartic term $ \lambda_{H \eta \phi} H^{\dagger} \eta \phi^*_1 \phi_2$ in the Lagrangian \eqref{scalar:lag}. This can be realised by minimising the scalar potential with respect to $\eta$. This leads to 
\begin{equation}
\langle \eta'_R \rangle = v_{\nu} \approx \frac{\mu_{H\eta}  v u_1/\sqrt{2}+  \lambda_{H \eta \phi} v u_1 u_2/2}{2 \mu^2_{\eta}}
\label{etavev}
\end{equation}
To simplify the calculation we have assumed all two VEVs of singlet scalars are equal, i.e. $u_1=u_2=u$ and also assumed the induced VEV to be negligible. The mass of the new gauge boson after spontaneous symmetry breaking is 
\begin{equation}
 M_{Z_{BL}} = \sqrt{5} g_{BL} u
\end{equation}
where we have ignored the contribution due to $v_{\nu}$ as it is negligible compared to that from $u$.

After putting equation \eqref{H&phi_broken_phsae} in equation \eqref{scalar:lag} we have found out the 4$\times$4 mixing matrix for the real scalar fields in the basis $\frac{1}{\sqrt{2}}\left(h^\prime\,\, s^\prime_1\,\,s^\prime_2 \,\,\eta^\prime_R\right)^T$ which has the following form,
{\footnotesize
\begin{eqnarray}
\left(\rm 
\begin{array}{cccc}
2 v^2 \lambda_H & u\, v \lambda_{H \phi_1} & u\, v \lambda _{H\phi_2} & 0 \\
 u\, v \lambda_{H\phi_1} & 2 u^2 \lambda_{\phi_1} & u \left(u \lambda_\phi +\mu_\phi  \sqrt{2}\right) & 0 \\
 u\, v \lambda_{H\phi_2} & u \left(u \lambda_\phi +\mu_\phi  \sqrt{2}\right) & \frac{1}{2} u \left(4 u \lambda_{\phi_2}-\sqrt{2} \mu_\phi \right) & \frac{u\, v \lambda_{H\eta\phi }}{2} \\
 0 & 0 & \frac{u\, v \lambda_{H\eta\phi}}{2} & \frac{1}{2} \left(\lambda_{\eta \phi_1}+\lambda_{\eta \phi_2}) u^2+v^2 (\lambda_{H\eta}+\lambda_{H\eta}^\prime)+2 \mu_{\eta}^2\right) \\
\end{array}
\right)
\label{rl:sclr:mass}
\end{eqnarray}
}
The physical scalars $\frac{1}{\sqrt{2}}\left(h\,\, s_1\,\,s_2 \,\,\eta_R\right)$ can be obtained by diagonalising this real symmetric mass matrix and that can be done by the orthogonal matrix $\mathcal{O}_S$ and the physical states can be expressed as 

{\small{\begin{eqnarray}
\left(
\begin{array}{cccc}
 h \\
 s_1 \\
 s_2 \\
\eta_R \\
\end{array}
\right)=
\mathcal{O_S}^T
\left(
\begin{array}{cccc}
 h^\prime \\
 s_1^\prime \\
 s_2^\prime \\
\eta_R^\prime \\
\end{array}
\right)\,,
\end{eqnarray}}}
In a similar manner, the 3$\times$3 pseudo scalar mass matrix can be written as 
\begin{eqnarray}
\left(
\begin{array}{ccc}
 -2 \sqrt{2} u\, \mu_\phi  & \sqrt{2} u\, \mu_\phi  & u\,v \lambda_{H\eta \phi} \\
 \sqrt{2} u\, \mu_\phi  & -\frac{u \mu_\phi }{\sqrt{2}} & -\frac{1}{2} u\, v\, \lambda_{H\eta \phi} \\
 u\, v\, \lambda_{H \eta \phi } & -\frac{1}{2} u\, v \lambda_{H\eta \phi } & \frac{1}{2} \left((\lambda_{ \eta \phi_1}+\lambda_{\eta \phi_2}) u^2+v^2 (\lambda_{ H\eta}+\lambda_{H\eta}^\prime)+2 \mu_{\eta}^2\right) \\
\end{array}
\right)
\label{ps:scl:mass}
\end{eqnarray}

The physical pseudo-scalars and the Goldstone boson $\frac{1}{\sqrt{2}}\left(A_1\,\, A_2\,\,\eta_I\right)$ can be obtained by diagonalising the above mass matrix and that can be done by the orthogonal matrix $\mathcal{O_P}$ and the states can be expressed as 
{\small{\begin{eqnarray}
\left(
\begin{array}{ccc}
 A_1 \\
 A_2 \\
\eta_I \\
\end{array}
\right)=
\mathcal{O_P}^T
\left(
\begin{array}{ccc}
 A_1^\prime \\
 A_2^\prime \\
\eta_I^\prime \\
\end{array}
\right)\,,
\end{eqnarray}}}
After the analysing of the scalar potential and diagonalising the mass matrices there will be four independent quartic couplings left (${\rm \lambda_{\eta} \,\, , \lambda_{H\eta}}\,\, , \lambda_{\eta \phi_1}\, \, ,\lambda_{\eta \phi_2} $). All the other couplings in the potential can be expressed in terms of the VEVs, scalar masses and the mixing angles.  In principle there should be nine different mixing angles ($\sin \theta_{ij}$) present in the scalar sector out of which six will come from the real sector and three will come from the pseudo scalar sector. Later we have shown that the result in the DM sector is almost independent of these mixing angles ($\sin \theta_{ij}$) and through our discussion to simplify the numerical analysis we have assumed all of them to be equal to 0.1. The parametrisation of the orthogonal matrices $\mathcal{O_S}, \mathcal{O_P}$ are shown in appendix \ref{appendix1} and \ref{appendix2}.

Lets discuss the fermionic sector of our model. We have three  generations of right handed neutrinos and four chiral fermions and the corresponding interactions can be written as 
\begin{align}
\mathcal{L}_{fermion} = \mathcal{L}_{\nu_R} + \mathcal{L}_{\rm DM}
\end{align}
where $\mathcal{L}_{\nu_R}$ is the interactions related to the right handed neutrinos can be expressed as 
\begin{align}
\mathcal{L}_{\nu_R} &= i\,  \overline{{\nu}_{Rj}}
\slashed{D}(Q^R_{\nu}){\nu}_{Rj} + \left( Y_{ij} (\overline{\ell_L})_i i\tau_2 \eta^* \nu_{Rj} + \text{h.c.} \right)
\label{Lag:nuR}
\end{align}
The first term in the equation \eqref{Lag:nuR} represents the kinetic part of ${\rm \nu_R}$ and the second term is the Yukawa interaction between SM lepton doublet ${\rm \ell_L\, \, , \nu_R\, \, , \text{and}\,  \eta}$ which is responsible for generating neutrino mass. As discussed above, the neutral component of ${\rm \eta}$ will get a small induced VEV $v_{\nu}$ through the trilinear interaction present in the potential. This will generate a tiny Dirac neutrino mass as
\begin{equation}
(m_{\nu})_{ij} = \frac{Y_{ij} v_{\nu}}{\sqrt{2}}
\label{numass}
\end{equation}
As can be seen from equation \eqref{etavev}, a tiny induced VEV $v_{\nu} \approx \mathcal{O}(\rm eV)$ can be generated by appropriate tuning of the trilinear coupling $\mu_{H\eta} $, quartic coupling $\lambda_{H \eta \phi}$ as well as bare mass squared term $\mu^2_{\eta}$. Since $u \sim 10$ TeV, $v\sim 100$ GeV, we can have $v_{\nu} \sim 0.1$ eV by choosing $\mu_{H\eta}/\mu^2_{\eta} \sim 10^{-16} \; {\rm GeV}^{-1}$ and $\lambda_{H \eta \phi} /\mu^2_{\eta} \sim 10^{-20} \; {\rm GeV}^{-2}$ which can be ensured by choosing very large $\mu^2_{\eta}$. This also ensures that the components of $\eta$ decouple from the low energy particle spectra as well as their relevant phenomenology. The hierarchy between $\mu_{H \eta}$ and $\mu_{\eta}$ can be reduced to bring the ratio to $\mu_{H\eta}/\mu^2_{\eta} \sim 10^{-11} \; {\rm GeV}^{-1}$ if we tune the Dirac Yukawa couplings to be as small as electron Yukawa coupling. Without any fine-tuning of parameters, we consider $\mu_{H \eta} \sim u, \lambda_{H \eta \phi} \sim \mathcal{O}(1)$ so that $\mu_{\eta}$ is required to be very large $(\geq 10^{10} \; {\rm GeV})$ thereby decoupling the neutrinophillic scalar doublet $\eta$ from low energy spectrum. Fine tuning of these parameters will enable the scalar doublet $\eta$ to have lighter mass having consequences at colliders as well as for thermalisation of right handed part of light Dirac neutrinos. However, we do not pursue such aspects in our studies. Similar way of generating sub-eV Dirac neutrino mass from induced VEV of neutrinophilic Higgs was proposed earlier by the authors of \cite{Davidson:2009ha, Davidson:2010sf, Baek:2019wdn}.

The term $\mathcal{L}_{\rm DM}$ is the interactions correspond to the chiral fermions can be written as 
\begin{eqnarray} \nonumber
\mathcal{L}_{\rm DM} &=& i [\overline{{\chi}_L}
\slashed{D}(Q^L_{\chi}){\chi}_L 
+ \overline{{\chi}_R}
\slashed{D}(Q^R_{\chi}){\chi}_R + \overline{{\psi}_L}
\slashed{D}(Q^L_{\psi}){\psi}_L 
+ \overline{{\psi}_R}
\slashed{D}(Q^R_{\psi}){\psi}_R]- \\
&&\bigg(f_1\,\overline{{\chi_L}} {\chi}_R\,\phi^{*}_1
+ f_2\,\overline{{\psi}_L} {\psi}_R\,\phi_2^{*} + \text{h.c.} \bigg) \,.\,
\label{Lag:DM}
\end{eqnarray}

We now rewrite the above Lagrangian in the basis $\xi_1 = \chi_L+\chi_R$ and $\xi_2 = \psi_L+\psi_R$.
In the basis of $\xi_1$ and $\xi_2$, the above Lagrangian \eqref{Lag:DM} can be written as
\begin{eqnarray}
\mathcal{L}_{\rm DM} &=& i\,\overline{\xi_1}\,\slashed{\partial}\xi_1
+ i\,\overline{\xi_2}\,\slashed{\partial}\xi_2
-g_{BL}\left(\dfrac{13}{9}\right)\,\overline{\xi_1}\,\slashed{Z}_{BL}\,P_L\,\xi_1\,
-g_{BL}\left(\dfrac{1}{9}\right)\,\overline{\xi_2}\,\slashed{Z}_{BL}\,P_L\,\xi_2\, \nonumber \\
&&-g_{BL}\left(\dfrac{22}{9}\right)\,\overline{\xi_1}\,\slashed{Z}_{BL}\,P_R\,\xi_1\,
-g_{BL}\left(\dfrac{19}{9}\right)\,\overline{\xi_2}\,\slashed{Z}_{BL}\,P_R\,\xi_2\,
-f_1\,\overline{\xi_1}\,P_R\,{\xi_1}\,\phi^{\dagger}_1 \nonumber \\
&&
- f_2\,\overline{\xi_2}\,P_R\,{\xi_2}\,\phi_2
-f_1\,\overline{\xi_1}\,P_L\,{\xi_1}\,\phi_1
- f_2\,\overline{\xi_2}\,P_L\,{\xi_2}\,\phi^\dagger_2 
\,, 
\label{Lag:DM:simp}
\end{eqnarray}
where $P_{L,R} = \dfrac{1 \pm \gamma_5}{2}$, left and right chiral projection
operators. From the above Lagrangian \eqref{Lag:DM:simp} it is clear that DM particles will get mass after the breaking of $B-L$ symmetry spontaneously by the VEV's of the singlet scalars ($\phi$s). $\xi_1$ and $\xi_2$ can annihilate to the SM particles through the interaction with $Z_{BL}$ and the singlet scalars. Due to the suitable choice of the scalar sector of the model, in the basis $\xi_1 = \chi_L+\chi_R$ and $\xi_2 = \psi_L+\psi_R$, all the interactions in equation \eqref{Lag:DM:simp} are exactly diagonal in $(\xi_1, \xi_2)$ basis. This is similar to imposing two different ${\rm Z_2}$ symmetries to two different DM candidates as: $Z_2: \xi_1 \rightarrow -\xi_1, Z^{\prime}_2: \xi_2 \rightarrow -\xi_2$ while all other particles being even under these symmetries. Clearly, the complete Lagrangian of our model is invariant under these two remnant discrete symmetries. Therefore, $\xi_1$ and $\xi_2$ are completely stable and will play the roles of two dark matter candidates in this model. 

\section{Constraints on the model parameters}
\label{sec:constraints}
Before discussing our results, we first note down the existing constraints on the model parameters from both theory and experiments. We discuss them one by one in this section as follows.

\subsection{Boundedness of Scalar Potential}
The scalar potential of the model has to be bounded from below and that can be ensured by the following inequalities.
\begin{eqnarray}\nonumber
\lambda_{H},\lambda_{\eta},\lambda_{\phi_1},\lambda_{\phi_2},\lambda^{\prime}_{H\eta},\lambda_{H\eta\phi} \geq 0\,,\\ \nonumber
\lambda_{H\phi_2}+\sqrt{\lambda_{H} \lambda_{\phi_2}}\geq 0\,,
\lambda_{H\phi_1}+\sqrt{\lambda_{H} \lambda_{\phi_1}}\geq 0\,,\\ 
\lambda_{H\eta}+\sqrt{\lambda_{H} \lambda_{\eta}}\geq 0\,\, ,
\lambda_{\phi}+\sqrt{\lambda_{\phi_1} \lambda_{\phi_2}}\geq 0\,,\\ \nonumber
\lambda_{\eta\phi_1}+\sqrt{\lambda_{\eta} \lambda_{\phi_1}}\geq 0\,,
\lambda_{\eta\phi_2}+\sqrt{\lambda_{\eta} \lambda_{\phi_2}}\geq 0\,.
\label{bounded}
\end{eqnarray}

\subsection{Perturbativity of Couplings}

We have to also take care of the perturbative breakdown of the model and to to guarantee that all quartic, Yukawa and gauge couplings should obey the following conditions.
\begin{eqnarray}
|\lambda_H| < 4 \pi,~|\lambda_{\phi_{1,2}}| < 4 \pi,~|\lambda_{\eta}| < 4 \pi,~\nonumber\\
~|\lambda_{H\phi_{1,2}}| < 4 \pi,
|\lambda_{\eta \phi_{1,2}}| < 4 \pi,~|\lambda_{\phi}| < 4 \pi,~\nonumber\\
~|\lambda_{H \eta}| < 4 \pi,~|\lambda^{\prime}_{H\eta}| < 4 \pi,~|\lambda_{H\eta\phi}| < 4 \pi,~\nonumber\\
|f_i| < \sqrt{4 \pi},~ |Y_{i,j}| < \sqrt{4 \pi},~\nonumber \\
|g, g'| < \sqrt{4\pi},~|g_{BL}| < \sqrt{4\pi} ,
\label{eq:PerC}
\end{eqnarray}

\subsection{Collider Constraints}
Apart from the theoretical constraints mentioned above, there exists stringent experimental constraints on the $B-L$ gauge sector. The limits from LEP II data constrains such additional gauge sector by imposing a lower bound on the ratio of new gauge boson mass to the new gauge coupling $M_{Z'}/g' \geq 7$ TeV \cite{Carena:2004xs, Cacciapaglia:2006pk}. The bounds from ongoing LHC experiment have already surpassed the LEP II bounds. In particular, search for high mass dilepton resonances have put strict bounds on such additional gauge sector coupling to all generations of leptons and quarks with coupling similar to electroweak ones. The latest bounds from the ATLAS experiment \cite{Aaboud:2017buh, Aad:2019fac} and the CMS experiment \cite{Sirunyan:2018exx} at the LHC rule out such gauge boson masses below 4-5 TeV from analysis of 13 TeV data. Such bounds get weaker, if the corresponding gauge couplings are weaker \cite{Aaboud:2017buh} than the electroweak gauge couplings. Also, if the $Z'$ gauge boson couples only to the third generation of leptons, all such collider bounds become much weaker, as explored in the context of DM and collider searches in a recent work \cite{Barman:2019aku}. Apart from the additional gauge boson, the additional singlet scalar spectrum is also constrained by experimental data. Though the singlet scalars do not directly couple to the SM particles, they can do so by virtue of their mixing with the SM Higgs. Such singlet scalar - Higgs mixing faces both theoretical and experimental constraints \cite{Robens:2015gla,Chalons:2016jeu}.  In case of scalar singlet extension of SM, the strongest bound on scalar-SM Higgs mixing angle ($\theta_{1j}, j=2,3,4$) comes form $W$ boson mass correction \cite{Lopez-Val:2014jva} at NLO for $250 {\rm ~ GeV} \lesssim M_{s_i} \lesssim 850$ GeV as ($0.2 \lesssim \sin\theta_{1j} \lesssim 0.3$) where $M_{s_i}$ is the mass of other physical Higgs. Whereas, for $M_{s_i}>850$ GeV, the bounds from the requirement of perturbativity and unitarity of the theory turn dominant  which gives $\sin\theta_{1j}\lesssim 0.2$. For lower values {\it i.e.} $M_{s_i}<250$ GeV, the LHC and LEP direct search \cite{Khachatryan:2015cwa,Strassler:2006ri} and measured Higgs signal strength \cite{Strassler:2006ri} restrict the mixing angle $\sin\theta_{1j}$ dominantly ($\lesssim 0.25$). The bounds from the measured value of EW precision parameter are mild for $M_{s_i}< 1$ TeV. While these constraints restrict the singlet scalar mixing with SM Higgs denoted by ($\theta_{1j}, j=2,3,4$), the other three angles ($\theta_{23}, \theta_{24}, \theta_{34}$) remain unconstrained. We choose our benchmark values of singlet scalar masses and their mixing with SM Higgs boson in such a way that these constraints are automatically satisfied.

\subsection{Cosmological Bound on Additional Light Degrees of Freedom}
Another interesting way to constrain the model parameters is by calculating the additional relativistic degrees of freedom due to the presence of right handed neutrinos at sub-eV scale having sizeable gauge interactions. Through these gauge interactions, they will achieve the thermal equilibrium in the early universe and will contribute to the total relativistic degrees of freedom of the thermal plasma. However, the total effective degrees of freedom for neutrinos are already very much constrained from cosmological observations, more specifically from BBN and CMB. We have used this fact to constrain the parameter space of the model. Recent data from the CMB measurement by the Planck\cite{Aghanim:2018eyx} suggests that the effective degrees of freedom for neutrinos as 
\begin{eqnarray}
{\rm
N_{eff}= 2.99^{+0.34}_{-0.33}
}
\label{Neff}
\end{eqnarray}
In this scenario the effective contribution from the right-handed neutrinos can be written as \cite{Abazajian:2019oqj, FileviezPerez:2019cyn}
\begin{eqnarray}
{\rm
\Delta N_{eff}=N_{eff}-N^{SM}_{eff}= N_{\nu_R}\left(\frac{T_{\nu_R}}{T_{\nu_L}}\right)^4= N_{\nu_R}\left(\frac{g\left(T^{dec}_{\nu_L}\right)}{g\left(T^{dec}_{\nu_R}\right)}\right)^{4/3}
}
\label{decoup0}
\end{eqnarray}
where $\rm{N_{\nu_R}}$ represents the number of relativistic right-handed neutrinos, g(T) corresponds to the relativistic degrees of freedom at temperature T, and ${\rm T^{dec}_{\nu_R}\, \, , T^{dec}_{\nu_L}}$ are the decoupling temperatures for ${\rm \nu_R}$ and ${\rm \nu_L}$ respectively. From equation \eqref{Neff} one can write 
\begin{equation}
{\rm \Delta N_{eff}=N_{eff}-N^{SM}_{eff}\leq 0.285}
\label{DeltaN}
\end{equation}
where we have used ${\rm N^{SM}_{eff}}=3.045$\cite{deSalas:2016ztq} and considered the maximum allowed ${\rm N_{eff}}$ from CMB bound mentioned above in equation \eqref{Neff}. Now, to predict ${\rm \Delta N_{eff}}$ one needs to know the decoupling temperature of ${\rm \nu_R}$ which remains in thermal equilibrium until the interaction rate becomes smaller than the Hubble expansion of the universe. 
\begin{equation}
{\rm 
\Gamma_{\nu_R} (T^{dec}_{\nu_R})= H(T^{dec}_{\nu_R})
}
\label{decoup}
\end{equation}
Here the Hubble rate can be written as \cite{FileviezPerez:2019cyn}
\begin{eqnarray}
H(T) = \sqrt{\frac{8 \pi G_N \rho(T)}{3}}=\sqrt{\frac{4\pi^3 G_N}{45} \left( g(T) +   \displaystyle 3\frac{7}{8} g_{\nu_R} \right)}\, \,T^2,
\end{eqnarray}
where $g_{\nu_R}$ is the internal degrees of freedom for right-handed neutrinos. In this scenario, the interaction rate can be written as\cite{FileviezPerez:2019cyn}
\begin{eqnarray}
\Gamma_{\nu_R} (T) &=& n_{\nu_R}(T) \langle \sigma (\bar{\nu}_R \nu_R \to \bar{f} f ) \, v_M \rangle \\
&=&\frac{g_{\nu_R}^2}{n_{\nu_R}(T)}\int \frac{d^3\vec{p}}{(2\pi)^3} f_{\nu_R}(p) \int \frac{d^3\vec{k}}{(2\pi)^3} f_{\nu_R}(k) \sigma_f(s) v_M, \nonumber 
\end{eqnarray} 
where ${\rm f(\nu_R)}$ is the Fermi-Dirac distribution of right-handed neutrinos. As we have discussed earlier, ${\rm \nu_R}$ will achieve thermal equilibrium only through ${\rm Z_{BL}}$ interactions and the cross-section can be written as
\begin{equation}
\sigma_{ \bar{\nu}_R\nu_R\to \bar{f}f } = \frac{g_{BL}^4}{12 \pi \sqrt{s}}\frac{1}{(s-M_{Z_{BL}}^2)^2 + \Gamma_{Z_{BL}}^2M_{Z_{BL}}^2}\sum_f N_f^C  n_f^2\sqrt{s-4M_f^2}\, (2M_f^2+s),
\end{equation}
where ${\rm n_{f}}$ is is the charge of the SM fermions under ${\rm U(1)_{B-L}}$, ${\rm N^C_{f}}$ is the colour multiplicity of the fermions. Inserting the required input in equation \eqref{decoup} one can find out the decoupling temperature for right-handed neutrinos and using equations \eqref{decoup0}, \eqref{DeltaN} we can derive a bound on the unknown parameters of the model and in this case these are ${\rm g_{BL}\, \, \text{and} \, \, M_{Z_{BL}}}$.  In fact, this is not a feature of this model but can be applicable to any gauge symmetric model with additional light degrees of freedom having sizeable gauge interactions. For example, in left-right symmetric models with light Dirac neutrinos or light right handed neutrinos one can derive similar bounds on additional gauge sector, as discussed by several earlier works including \cite{Borah:2016lrl, Borah:2017leo} and references therein. It should be noted that the right handed neutrinos can also thermalise via Yukawa couplings $Y_{ij} (\overline{\ell_L})_i i\tau_2 \eta^* \nu_{Rj}$ depending upon its relative strength compared to gauge coupling. However, since we consider $\eta$ and its components to be very heavy and hence decoupled from low energy spectrum, the bounds on such Yukawa coupling from CMB-BBN constraints will be much weaker and hence we do not discuss it here.

\section{Dark Matter: Relic Density and Direct Detection}
\label{sec3}
Relic abundance of two component DM in our model $\chi_{1,2}$ can be found by numerically solving the corresponding Boltzmann equations. Let $n_2 = n_{\xi_2} + n_{\bar{\xi}_2}$ and
$n_1=n_{\xi_1} + n_{\bar{\xi}_1}$ are the total
number densities of two dark matter
candidates respectively. Assuming there
is no asymmetry in number densities of $\xi_i$
and $\bar{\xi}_i$, the two coupled Boltzmann
equations in terms of $n_2$ and $n_1$ are given below
\cite{Biswas:2019ygr},   
\begin{eqnarray}
\frac{dn_{2}}{dt} + 3n_{2} H &=& 
-\dfrac{1}{2}\langle{\sigma {\rm{v}}}_{\xi_2 \bar{\xi_2} \rightarrow {X \bar{X}}}\rangle 
\left(n_{2}^2 -(n_{2}^{\rm eq})^2\right)
- \dfrac{1}{2}{\langle{\sigma {\rm{v}}}_{\xi_2 \bar{\xi_2}
\rightarrow \xi_1 \bar{\xi_1}}\rangle} \bigg(n_{2}^2 - 
\frac{(n_{2}^{\rm eq})^2}{(n_{1}^{\rm eq})^2}n_{1}^2\bigg) \,,
%
\label{boltz-eq1} \\
\frac{dn_{1}}{dt} + 3n_{1} H &=& -\dfrac{1}{2}\langle{\sigma {\rm{v}}}
_{\xi_1 \bar{\xi_1} \rightarrow {X \bar{X}}}\rangle \left(n_{1}^2 -
(n_{1}^{\rm eq})^2\right) 
+ \dfrac{1}{2}{\langle{\sigma {\rm{v}}}_{\xi_2 \bar{\xi_2} \rightarrow {\xi_1} \bar{\xi_1}}\rangle} 
\bigg(n_{2}^2 - \frac{(n_{2}^{\rm eq})^2}{(n_{1}^{\rm eq})^2}
n_{1}^2\bigg)\,,
\label{boltz-eq2} 
\end{eqnarray}
where, $n^{\rm eq}_i$ is the equilibrium number density
of dark matter species $i$ and $H$ denotes the Hubble parameter, defined earlier. For further details of these Boltzmann equations for two component Dirac fermion DM and their annihilation channels ($\xi_i \bar{\xi_i} \rightarrow X \bar{X}$, X being all particles where DM can annihilate into)  contributing to $\langle{\sigma {\rm{v}}} \rangle$, please refer to \cite{Biswas:2019ygr} where a similar scenario was discussed recently. We have solved these two coupled Boltzmann equations using 
\texttt{micrOMEGAs} \cite{Belanger:2014vza} where the model information has
been supplied to \texttt{micrOMEGAs} using \texttt{FeynRules}
\cite{Alloul:2013bka}. All the relevant annihilation cross sections
of dark matter number changing  processes required to solve
the coupled equations are calculated using \texttt{CalcHEP} \cite{Belyaev:2012qa}. The most important DM annihilation channels are the ones mediated by $Z_{BL}$ and the singlet scalars. Since the two DM candidates are stabilised by two separate and accidental $Z_2$ symmetries, there is no coannihilation between them. On the other hand a pair of one DM can annihilate into a pair of the other, if kinematically allowed, as shown by the last terms on the right hand side of above two equations.

\begin{figure}[h!]
\centering
\includegraphics[height=3cm,width=5cm]{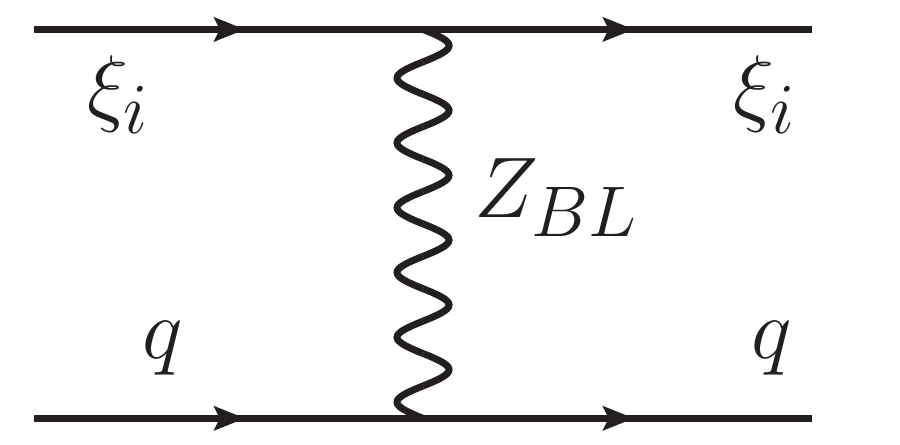}
\includegraphics[height=3cm,width=5cm]{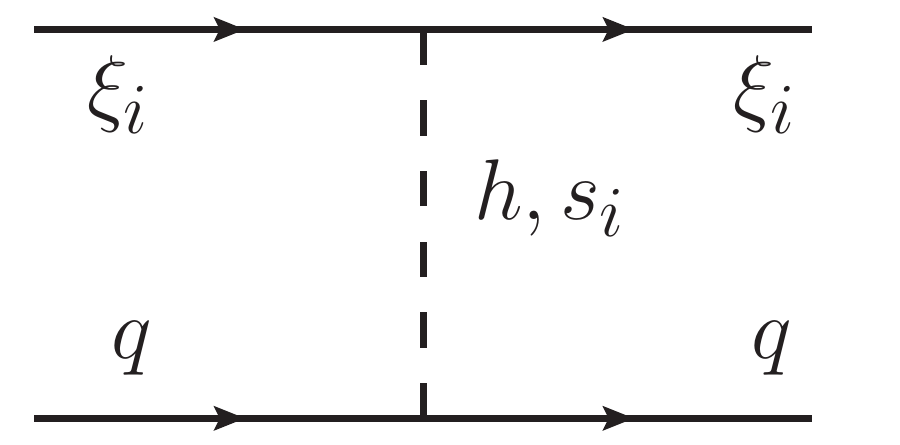}
\caption{Feynman diagrams for spin-independent elastic scattering processes of DM with nucleons (or quarks) in the model.}
\label{Fig:feyn_DD}
\end{figure}

Just like the new gauge boson and singlet scalars mediate DM annihilation into SM particles, similarly, they can also mediate spin independent DM-nucleon scatterings. The Feynman diagrams corresponding to such direct detection (DD) processes are shown in the figure \ref{Fig:feyn_DD}.
Different ongoing experiments like Xenon1T \cite{Aprile:2017iyp, Aprile:2018dbl}, LUX \cite{Akerib:2016vxi}, PandaX-II \cite{Tan:2016zwf, Cui:2017nnn} are trying to detect the DM in the lab-based experiments and give a strong upper bound on the spin-independent (SI) DD cross-section as a function of DM mass. We have extracted the SI elastic scattering cross-section for both the DM candidates from micrOmegas. DD analysis for two-component DM is slightly different from the single component scenario. To compare the result of our model with Xenon1T bound, we have multiplied the elastic scattering cross-section by the relative number density of each DM candidate and used the following conditions 

\begin{eqnarray}\nonumber
{\rm \sigma_{\xi_1}^{eff}=\frac{n_{\xi_1}}{n_{\xi_1}+n_{\xi_2}} \sigma_{\xi_1}^{SI}\leq \sigma_{Xenon1T} }\\  
{\rm \sigma_{\xi_2}^{eff}=\frac{n_{\xi_2}}{n_{\xi_1}+n_{\xi_2}} \sigma_{\xi_2}^{SI}\leq \sigma_{Xenon1T} }
\label{eff:DD}
\end{eqnarray} 

For details regarding direct detection of multi component DM, please refer to \cite{Herrero-Garcia:2017vrl, Herrero-Garcia:2018mky}.

\section{Results and Discussion}
\label{sec4}
\begin{figure}[h!]
\centering
\subfigure[]
{\includegraphics[scale=0.45]{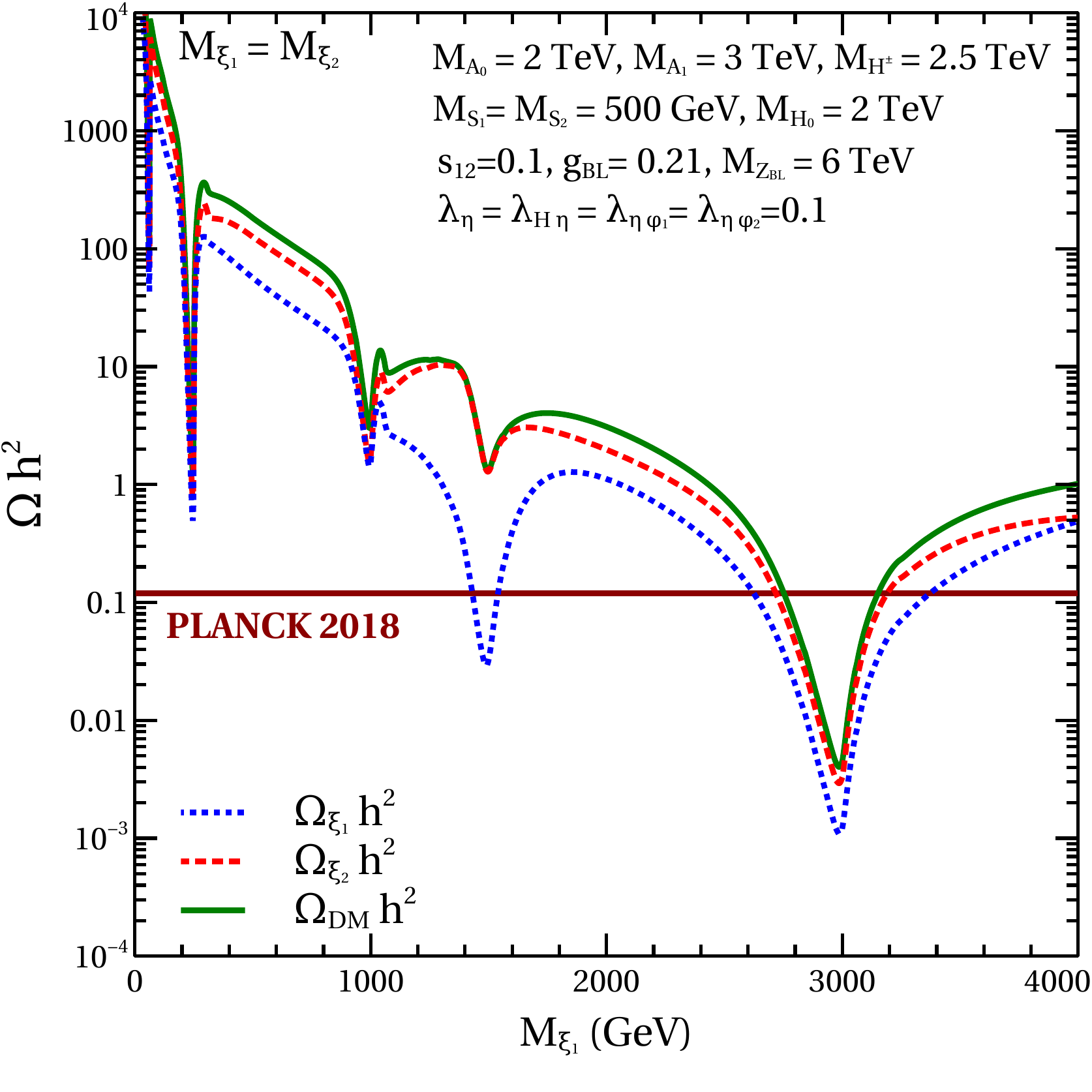}}
\,\,
\subfigure[]
{\includegraphics[scale=0.45]{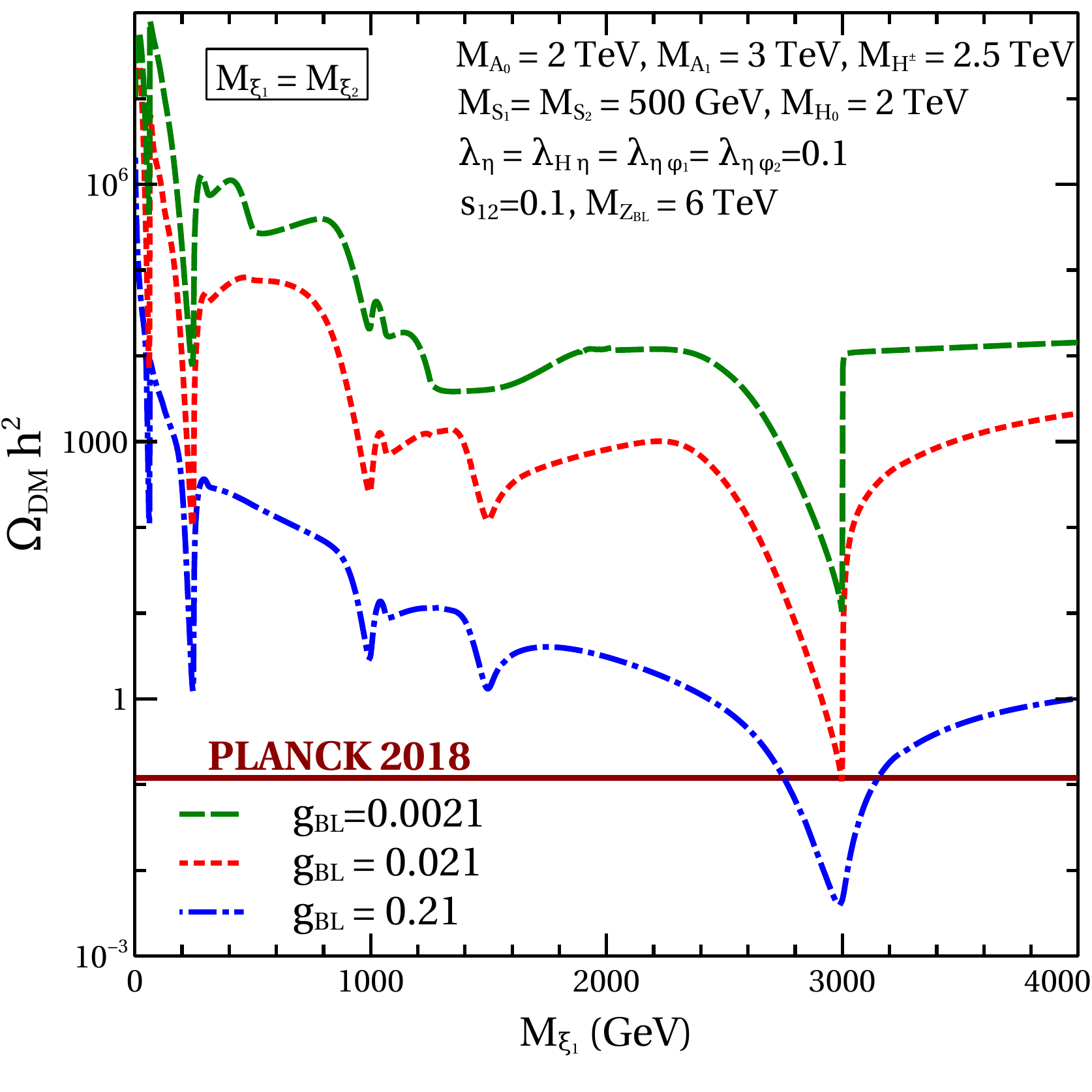}}
\caption{Left panel: Relic abundance of two DM candidates with degenerate masses keeping all other model parameters fixed to benchmark values. Right panel: Total relic abundance of two DM candidates $\left({\rm \Omega_{DM}h^2}= \Omega_{\xi_1}h^2  +  \Omega_{\xi_2}h^2\right)$ with degenerate masses ${\rm M_{\xi_1}=M_{\xi_2}}$ for different benchmark values of $U(1)_{B-L}$ gauge coupling.}
\label{Fig:omega-vs-m1-1}
\end{figure}

\begin{figure}[h!]
\centering
\subfigure[]
{\includegraphics[scale=0.45]{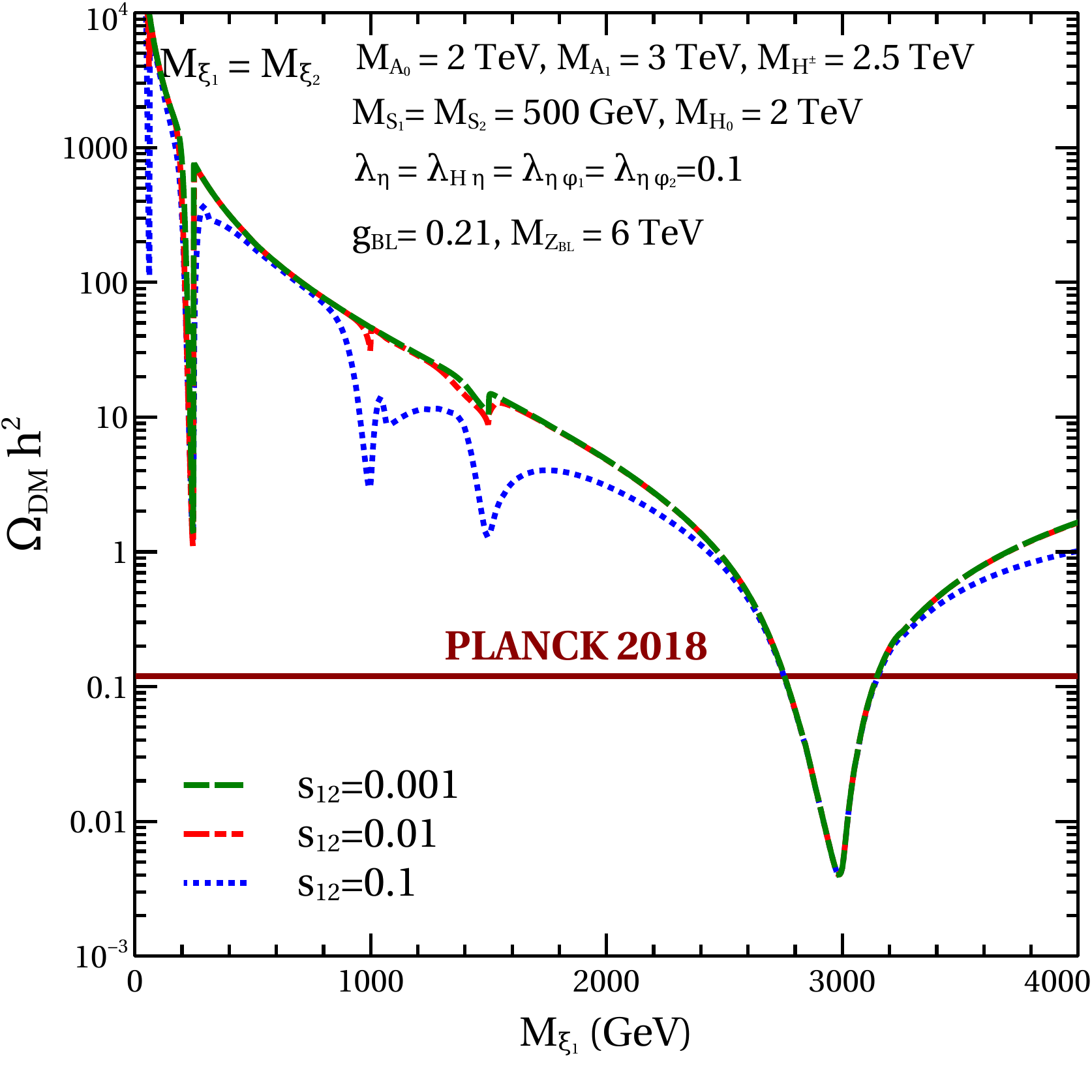}}
\,\,
\subfigure[]
{\includegraphics[scale=0.45]{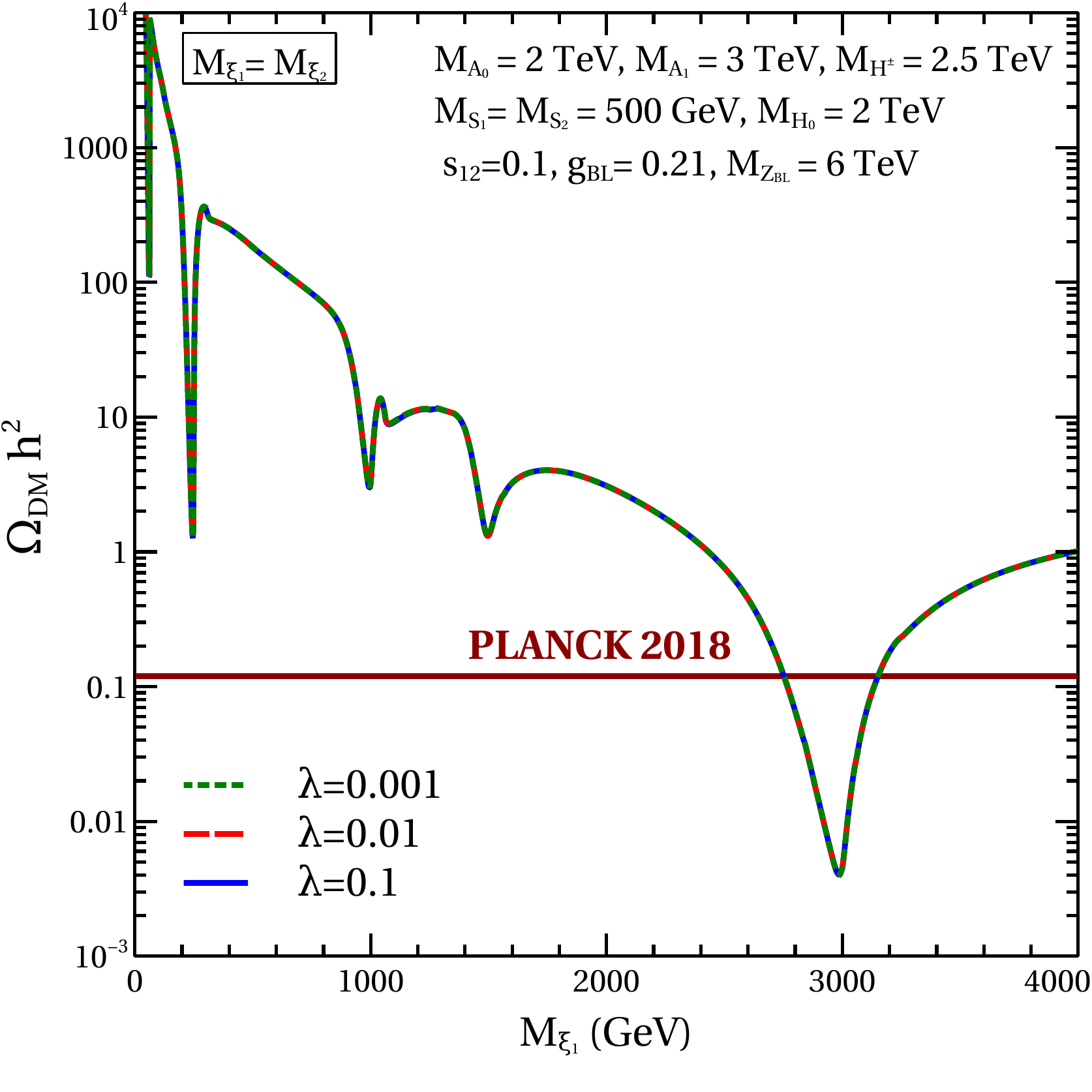}}
\caption{Total relic abundance of two DM candidates $\left({\rm \Omega_{DM}h^2}= \Omega_{\xi_1}h^2  +  \Omega_{\xi_2}h^2\right)$ with degenerate masses ${\rm M_{\xi_1}=M_{\xi_2}}$ for different benchmark values of: mixing angle (left panel) and (b) quartic couplings (right panel).}
\label{Fig:omega-vs-m1-2}
\end{figure}

Since we have two stable DM candidates i.e. $\xi_1$ and $\xi_2$ in this model, the total relic abundance can be expressed as the sum of the individual candidates, ${\rm\Omega_{DM} h^2 = \Omega_{\xi_1} h^2 + \Omega_{\xi_2} h^2}$. Equation \eqref{Lag:DM:simp} clearly shows that $\xi_1$ and $\xi_2$ have interactions with $Z_{BL}$ and the new singlet scalars $\phi_1$ and $\phi_2$. Through these interactions they will achieve the thermal equilibrium in the early universe (unless the gauge and Yukawa couplings are extremely small) and eventually freeze-out as the universe expands. In figure \ref{Fig:omega-vs-m1-1} and \ref{Fig:omega-vs-m1-2}, we have shown the dependence of relic abundance on DM mass by keeping the other parameters fixed at some benchmark values. For these two plots we assumed both the DM to have equal masses (${\rm M_{\xi_1} = M_{\xi_2}}$), although in principle, they can have different masses. The left panel of figure \ref{Fig:omega-vs-m1-1} shows the variation of relic abundance as a function of DM mass and the other parameters were chosen as $\rm {M_{\eta_R}=M_{\eta_I}=1.5\ TeV, M_{A_1}=2\ TeV, M_{\eta^\pm}=750\ GeV, M_{s_1}=M_{s_2} = \ 1\ TeV,}$ $\rm{\\ M_{Z_{BL}}=6\ TeV, g_{BL}=0.21}$ while all the scalar mixing angles $\left(\sin \theta_{ij}\right)$ and the independent quartic couplings are assumed to be equal to 0.1. Figure \ref{Fig:omega-vs-m1-1}a clearly shows the dip in the relic densities due to different scalars and ${\rm Z_{BL}}$ resonances, at DM mass of 62.5 GeV, 250 GeV, 1 TeV, 1.5 TeV, and 3 TeV respectively. The dotted blue line and the dashed red line represent the ${\rm \Omega_{\xi_1} h^2}$ and ${\rm \Omega_{\xi_2} h^2}$ respectively whereas the green solid line shows the total DM relic density. One important point to note here is that ${\rm \xi_2}$ has dominant contribution throughout the whole mass range and that is because of the ${B-L}$ charges assigned for the individual chiral fermions which constitute the two Dirac fermion DM candidates. In figure \ref{Fig:omega-vs-m1-1}b we have shown the dependence of total DM abundance on the gauge coupling ${\rm g_{BL}}$ which shows that the total relic abundance is decreasing as we are increasing the gauge coupling as expected.   

Figure \ref{Fig:omega-vs-m1-2} shows the dependence of DM abundance on the parameters from the scalar sector. The left panel of figure \ref{Fig:omega-vs-m1-2} is for different values of scalar mixing angle, (0.1, 0.01, 0.001) whereas the right one is for different quartic couplings (0.1,0.01,0.001). Both the figures clearly show that the total DM abundance does not have strong dependence on these two parameters.

\begin{figure}[h!!!]
\centering
\includegraphics[scale=0.5]{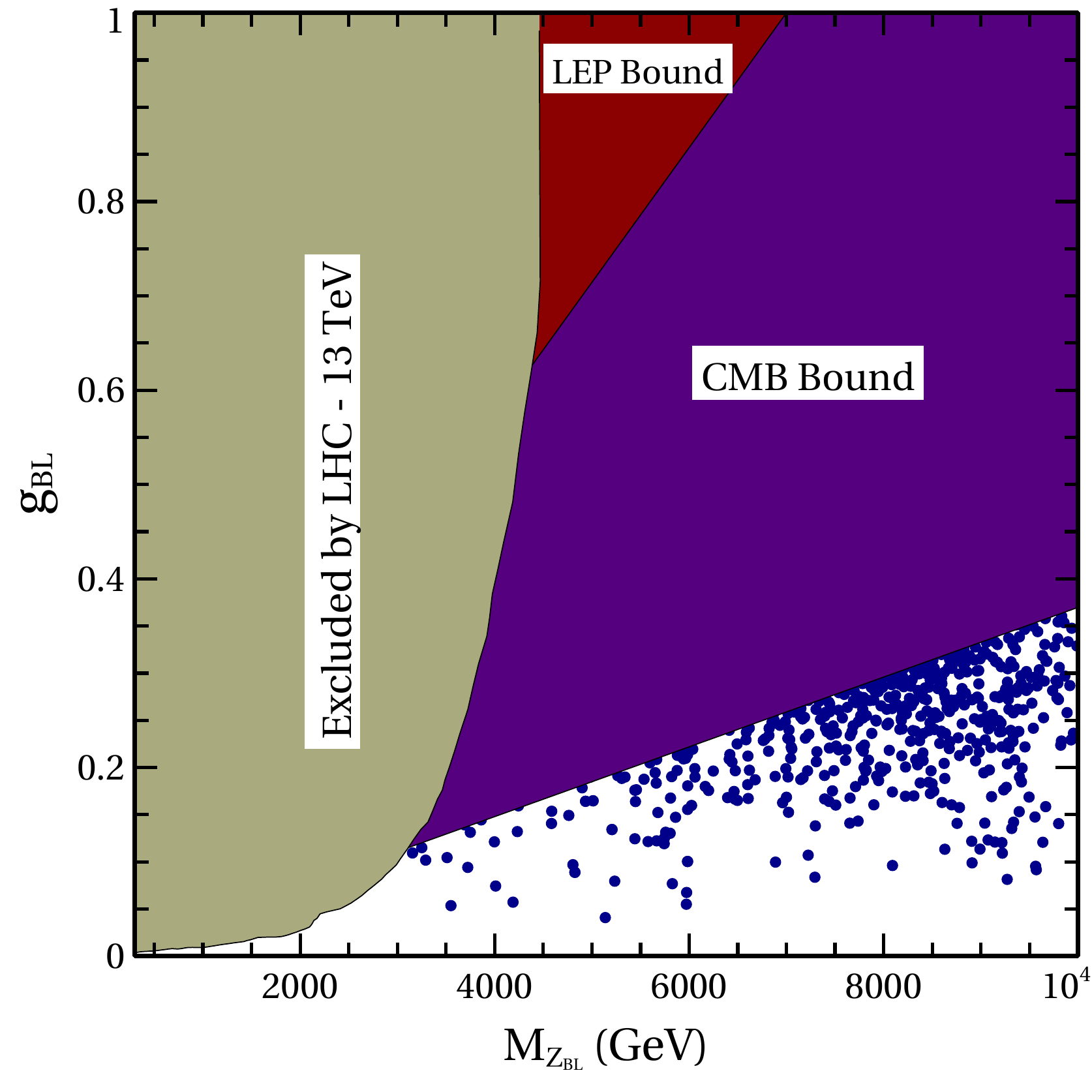}
\caption{Summary plot showing allowed parameter space of the model from all relevant experiments and observations. }
\label{scan-1}
\end{figure}

\begin{table}
\begin{center}
\begin{tabular}{|c|c|}
\hline
Parameters & Range   \\
\hline
\hline
$\rm{M_{\xi_1}}$ &  (10 GeV, 5 TeV)\\
$\rm{M_{\xi_2}}$ &  (10 GeV, 5 TeV)\\
\hline
$\rm{M_{Z_{BL}}}$ &  (100 GeV, 10 TeV) \\
$\rm{g_{BL}}$ &  (0.0001, 1)\\
\hline
$\rm{M_{s_1}}$ &  (100 GeV, 2 TeV)\\
$\rm{M_{s_2}}$ &  (100 GeV, 2 TeV)\\
$\rm{M_{A_1}}$ &  (100 GeV, 2 TeV)\\
\hline
$\rm{M_{\eta_R}}=\rm{M_{\eta_I}}$ & (1 TeV, 2 TeV)\\
$\rm{M_{\eta^\pm}}$ & 2.5 TeV\\
\hline
\end{tabular}
\end{center}
\caption{The parameters of the model and ranges used in the random scan}
\label{tab:scan}
\end{table} 
After analysing the dependence on different model parameters from the above benchmark plots we have now performed a random scan over the model parameters shown in the table \ref{tab:scan}. As mentioned earlier, we have kept the quartic couplings and the mixing angles fixed at 0.1 throughout our analysis unless otherwise specified. In figure \ref{scan-1} we have shown the final parameter space of this model in the ${\rm g_{BL}-M_{Z_{BL}}}$ plane where we have constrained the allowed parameter space from different relevant upper bounds coming from LHC, LEP, BBN-CMB and also XENON1T. The blue points showing in the above figure are allowed from all these experimental bounds. Apart from the experimental bounds, we also apply the bounded from below criteria of the scalar potential as well as perturbativity of all dimensionless couplings. One interesting point to note here is that the BBN-CMB bound on $\Delta N_{\rm eff}$ is putting much stronger bound in the high mass region of ${\rm M_{Z_{BL}}}$ compared to the other bounds like collider or direct detection. 

\begin{figure}[h!!!]
\centering
\subfigure[]
{\includegraphics[scale=0.45]{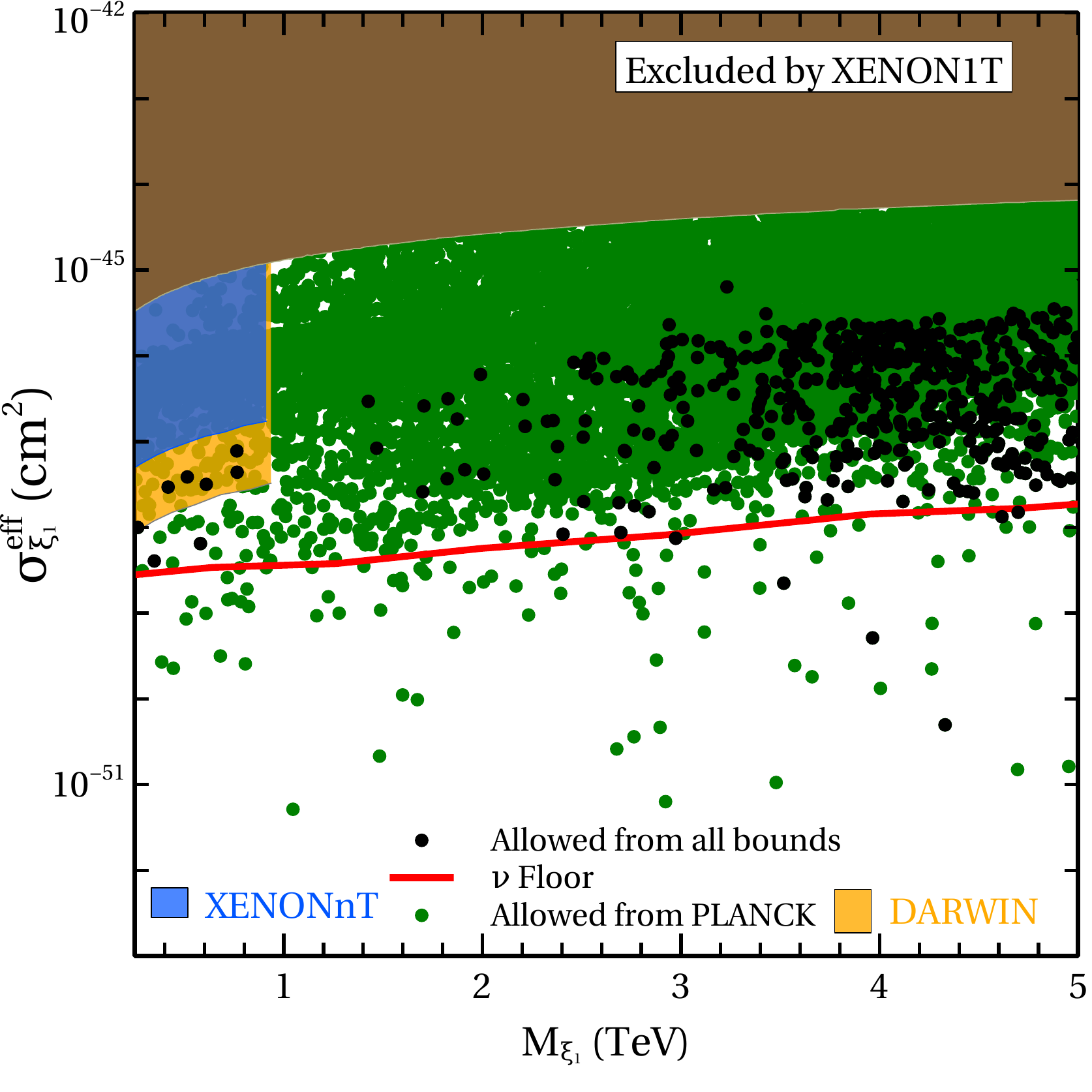}}
\,\,
\subfigure[]
{\includegraphics[scale=0.45]{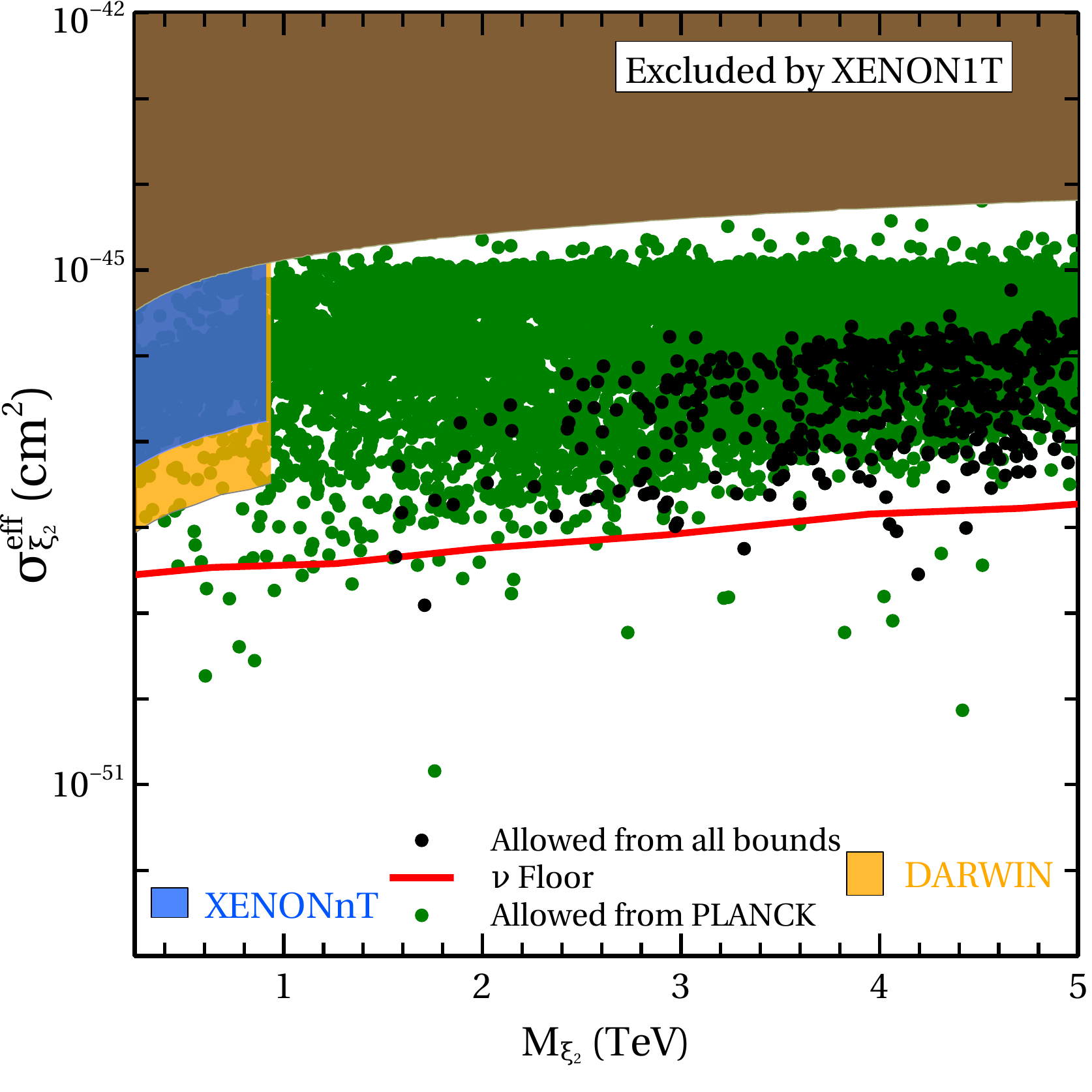}}
\caption{Effective spin-independent direct detection scattering cross-section of individual DM candidates. Green points are satisfying the boundedness of potential, perturbativity of couplings and the total DM relic density constraint whereas the black points are allowed from all relevant constraint such as direct detection, LHC, LEP and CMB bound on $\Delta N_{\rm eff}$. }
\label{Fig:DD}
\end{figure}

In order to show the prospects of probing such a scenario at ongoing and upcoming direct detection experiments, we have shown the effective spin-independent DD cross-section (see equation \eqref{eff:DD}) as function of individual DM mass in figure \ref{Fig:DD}. Points showed in green satisfy perturbativity of couplings, bounded from below criteria of the scalar potential, and the total DM relic density constraint whereas the black points are allowed from all other relevant constraints such as XENON1T, LHC, LEP and CMB bound on $\Delta N_{\rm eff}$ as well. From this figure, it is clear that the parameter space of this model has promising scope of being detected. We have also shown the projected sensitivities from future experiments such as XENONnT \cite{Aprile:2015uzo} (blue region) and DARWIN \cite{Aalbers:2016jon}  (yellow region) in both plots which clearly indicate that these two experiments can probe a large region of parameter space. For a comparison, we also show the neutrino floor by the red solid line, corresponding to coherent neutrino-nucleus scattering cross section \cite{Billard:2013qya}.

\begin{figure}[h!]
\centering
\subfigure[]
{\includegraphics[scale=0.45]{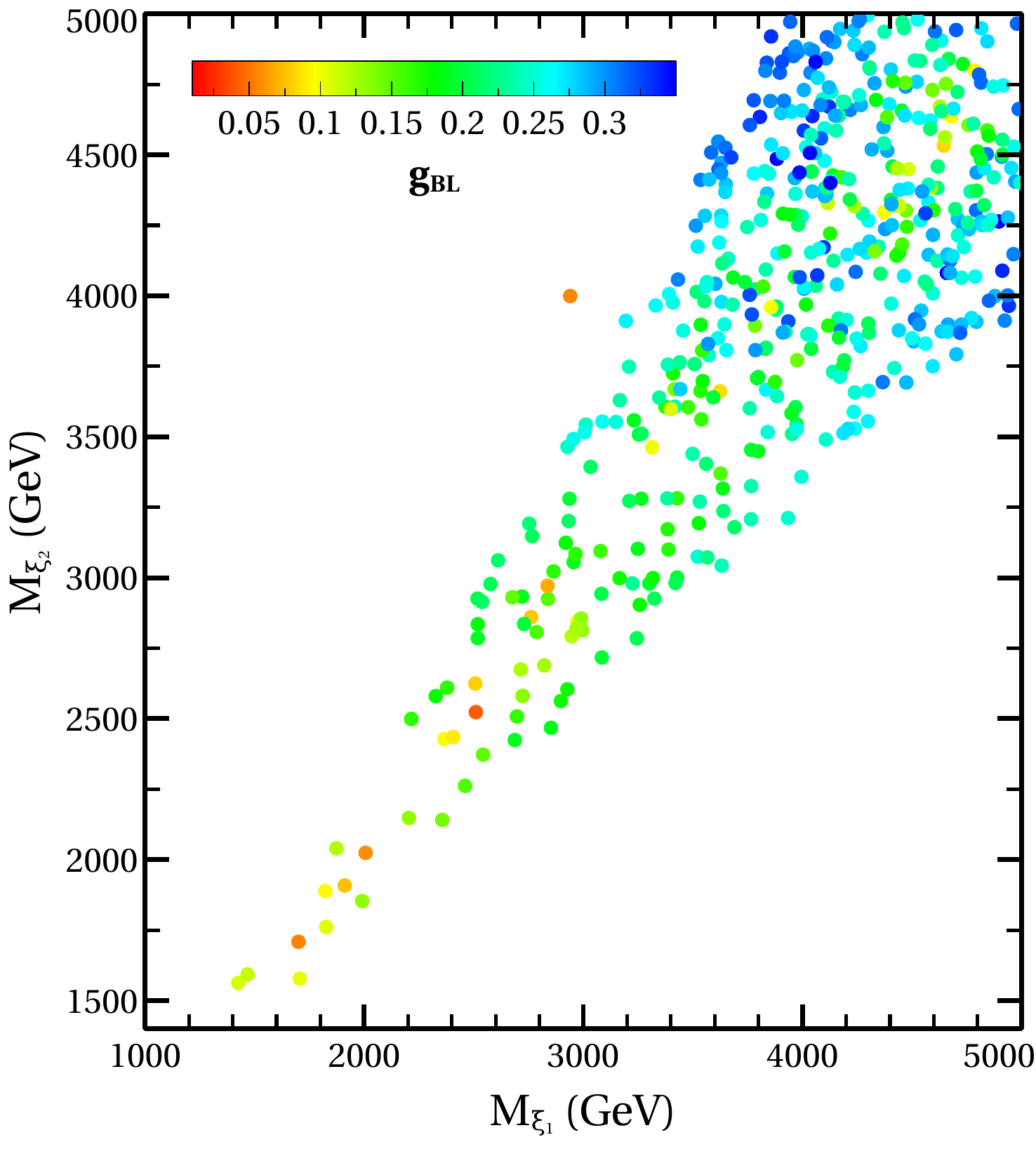}}
\,\,
\subfigure[]
{\includegraphics[scale=0.45]{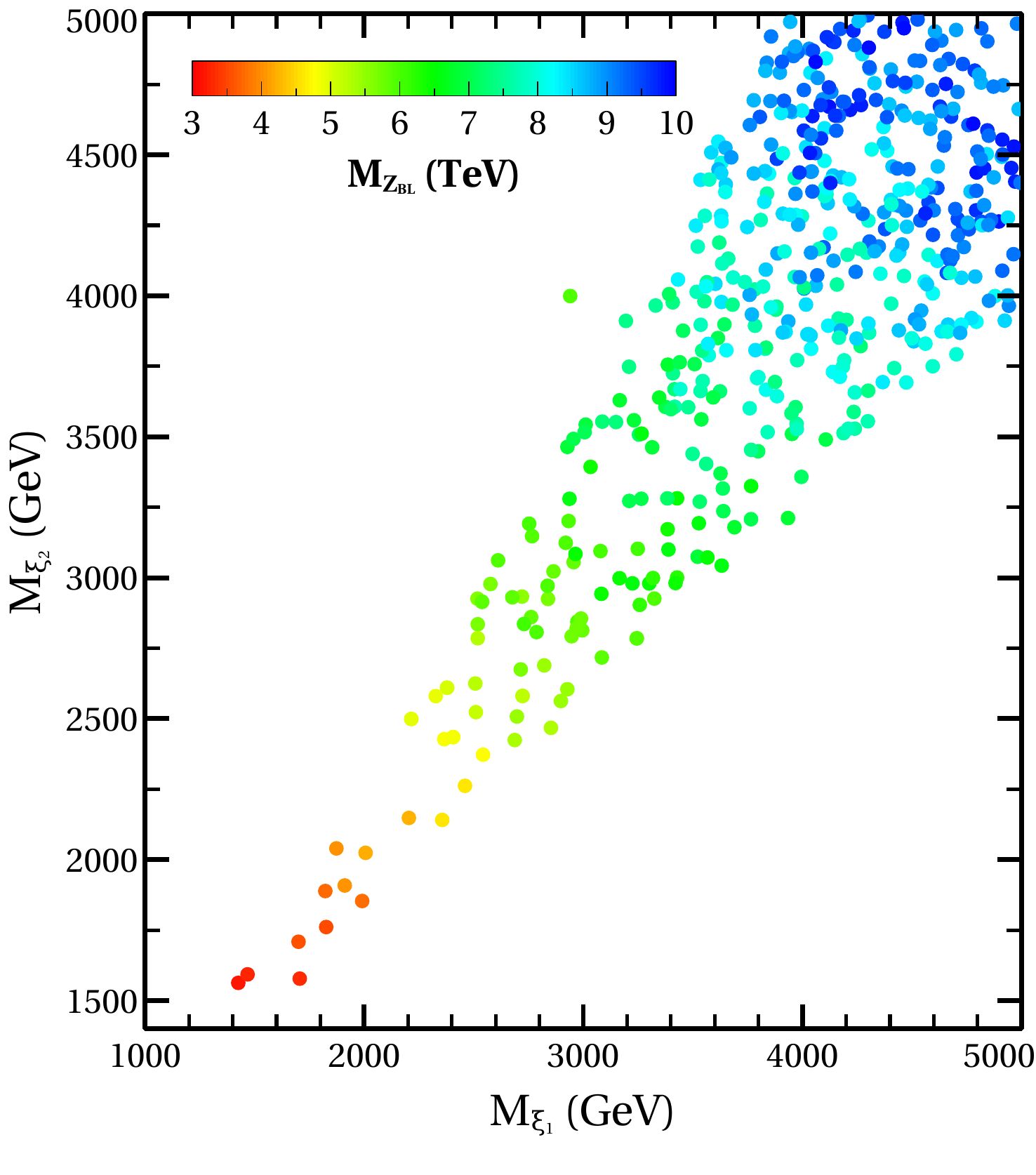}}
\caption{DM parameter in terms of DM masses space satisfying all relevant constraints. The colour coding is used to denote ${\rm g_{BL}}$, ${\rm M_{Z_{BL}}}$ in left and right panels respectively. }
\label{Fig:m1m2}
\end{figure}

In figure \ref{Fig:m1m2} we have shown the allowed parameter space in ${\rm M_{\xi_1}-M_{\xi_2}}$ plane where the variation of ${\rm g_{BL}}$ (right panel) and ${\rm M_{Z_{BL}}}$ (left panel) have also been shown through colour coding. One can see that the region becomes broader as we go to the high mass region of both DM candidates whereas in the low DM mass region it becomes narrower. This can be explained by noting the fact that the DM Yukawa couplings with singlet scalars namely, ${\rm f_i\,\, (i=1,2)}$ can be written as ${\rm \sqrt{2}\,\,M_{DM_i}/u}$ which will increase with ${\rm M_{DM_i}}$ for fixed $u$. So, in the low DM mass region, only gauge coupling (${\rm g_{BL}}$) is playing the major role in DM annihilation processes thereby deciding its relic. As discussed earlier, $B-L$ gauge boson portal interactions typically lead to correct DM relic around the resonance region ${\rm M_{DM_i}} \approx {\rm M_{Z_{BL}}}/2$. However, as the DM mass increases, Yukawa coupling corresponding to each DM candidate also increases and starts to contribute significantly taking the allowed parameter space away from the resonance region ${\rm M_{DM_i}} \approx {\rm M_{Z_{BL}}}/2$ mentioned before. As a result the region becomes broader as we go to the high mass region due to reduced dependence on the gauge boson mediated annihilation channels.

\section{Conclusion}
\label{sec5}
We have proposed a gauged $B-L$ model where light Dirac neutrinos and two component fermion DM can be realised without invoking the presence of additional discrete symmetries. The novel feature of this model is the possibility of sub-eV Dirac neutrino mass at tree level and existence of two stable fermion DM candidates within a minimal gauged $B-L$ symmetric framework. The right handed part of light Dirac neutrino and the two Dirac fermion DM candidates have appropriate $B-L$ charges to make the model anomaly free. The chosen $B-L$ charge of right handed part of light Dirac neutrino not only dictates the origin of Dirac neutrino mass from a neutrinophillic Higgs doublet with appropriate $B-L$ charge but also gives rise to the possibility of two stable fermion DM whose $B-L$ charges are related to that of right handed neutrinos via anomaly cancellation conditions. The scalar sector of the model is chosen in such a way that it not only gives rise to the desired particle masses but also leaves a remnant $Z_2 \times Z'_2$ symmetry after spontaneous symmetry breaking to stabilise the two DM candidates. We constrain the model parameters from different available constraints, both theoretical as well as experimental. After showing the behaviour of DM relic density for different choices of benchmark parameters, we perform a numerical scan to show the available parameter space in terms of DM masses and the parameters of the $B-L$ gauge sector namely, $g_{BL}, M_{Z_{BL}}$. Apart from the usual bounds from LEP, LHC, DM direct detection, DM relic density constraints, the strongest bound for $M_{Z_{BL}} >3$ TeV comes from BBN, CMB limits on the effective number of relativistic degrees of freedom. This interesting situation arises due to the Dirac nature of light neutrinos which introduces additional relativistic species (right handed part of light Dirac neutrinos) that can be thermalised in the early universe due to their sizeable $B-L$ gauge interactions. Thus, the model not only allows the possibility of sub-eV Dirac neutrino mass with two component fermion DM from the requirement of cancelling triangle anomalies but also leads to new contributions to $\Delta N_{\rm eff}$ that can be probed by future CMB experiments. The model also remains within reach of ongoing as well as near future direct detection experiments.  Thus the model not only offers a unified picture of light Dirac neutrino and multi-component DM but also allows the tantalising possibility of probing it at experiments operating at different frontiers.
\acknowledgements

DN would like to thank Arnab Dasgupta for useful discussions. DB acknowledges the support from Early Career Research Award from DST-SERB, Government of India (reference number: ECR/2017/001873). DB also thanks the Department of Physics, Tokyo Institute of Technology, Japan for kind hospitality during completion of some part of this work. 

\appendix
\section{The 4$\times$4 rotation matrices}
\label{appendix1}
\begin{eqnarray}
\mathcal{O_S}= \mathcal{O}_{S_{12}} \mathcal{O}_{S_{13}}\mathcal{O}_{S_{14}}\mathcal{O}_{S_{24}}\mathcal{O}_{S_{23}}\mathcal{O}_{S_{34}}
\end{eqnarray}
where $\mathcal{O}_{S_{ij}}$ represents the rotation in i-j plane.
\begin{eqnarray}
{\rm 
\mathcal{O}_{S_{12}}=\left(
\begin{array}{cccc}
 \cos\theta_{12} & \sin\theta_{12} & 0 & 0 \\
 -\sin\theta_{12} & \cos\theta_{12} & 0 & 0 \\
 0 & 0 & 1 & 0 \\
 0 & 0 & 0 & 1 \\
\end{array}
\right)\, ,
\mathcal{O}_{S_{13}}=\left(
\begin{array}{cccc}
 \cos\theta_{13} & 0 & \sin\theta_{13} & 0 \\
 0 & 1 & 0 & 0 \\
 -\sin\theta_{13} & 0 & \cos\theta_{13} & 0 \\
 0 & 0 & 0 & 1 \\
\end{array}
\right)
}
\end{eqnarray}

\begin{eqnarray}
{\rm
\mathcal{O}_{S_{14}}=\left(
\begin{array}{cccc}
 \cos\theta_{14} & 0 & 0 & \sin\theta_{14} \\
 0 & 1 & 0 & 0 \\
 0 & 0 & 1 & 0 \\
 -\sin\theta_{14} & 0 & 0 & \cos\theta_{14} \\
\end{array}
\right)\, ,
\mathcal{O}_{S_{24}}=\left(
\begin{array}{cccc}
 1 & 0 & 0 & 0 \\
 0 & \cos\theta_{24} & 0 & \sin\theta_{24} \\
 0 & 0 & 1 & 0 \\
 0 & -\sin\theta_{24} & 0 & \cos\theta_{24} \\
\end{array}
\right)
}
\end{eqnarray}

\begin{eqnarray}
{\rm
\mathcal{O}_{S_{23}}=\left(
\begin{array}{cccc}
 1 & 0 & 0 & 0 \\
 0 & \cos\theta_{23} & \sin\theta_{23} & 0 \\
 0 & -\sin\theta_{23} & \cos\theta_{23} & 0 \\
 0 & 0 & 0 & 1 \\
\end{array}
\right)\, ,
\mathcal{O}_{S_{34}}=\left(
\begin{array}{cccc}
 1 & 0 & 0 & 0 \\
 0 & 1 & 0 & 0 \\
 0 & 0 & \cos\theta_{34} & \sin\theta_{34} \\
 0 & 0 & -\sin\theta_{34} & \cos\theta_{34} \\
\end{array}
\right)
}
\end{eqnarray}

\section{The 3$\times$3 rotation matrices}
\label{appendix2}
\begin{eqnarray}
{\small
\mathcal{O_P}=\left(
\begin{array}{ccc}
 \cos\alpha_{12}\cos\alpha_{13}  & \sin\alpha_{12}\cos\alpha_{13} & \sin\alpha_{13}  \\
 -\sin\alpha_{12}\cos\alpha_{23}-\cos\alpha_{12}\sin\alpha_{23}\sin\alpha_{13} & \cos\alpha_{12}\cos\alpha_{23}-\sin\alpha_{12}\sin\alpha_{23}sin\alpha_{13} & \sin\alpha_{23}\cos\alpha_{13}  \\
 \sin\alpha_{12}\sin\alpha_{23}-\cos\alpha_{12}\cos\alpha_{23}\sin\alpha_{13} & -\cos\alpha_{12}\sin\alpha_{23}-\sin\alpha_{12}\cos\alpha_{23}\sin\alpha_{13} & \cos\alpha_{23}\cos\alpha_{13} \\
\end{array}
\right)
}
\end{eqnarray}
\providecommand{\href}[2]{#2}\begingroup\raggedright\endgroup

\end{document}